\documentclass[%
 aip,
 pof,
 amsmath,amssymb,
 preprint,%
]{revtex4-1}

\usepackage{graphicx}
\usepackage{dcolumn}
\usepackage{bm}

\usepackage[utf8]{inputenc}
\usepackage[T1]{fontenc}
\usepackage{mathptmx}
\usepackage{etoolbox}
\usepackage{url}

\makeatletter
\def\@email#1#2{%
 \endgroup
 \patchcmd{\titleblock@produce}
  {\frontmatter@RRAPformat}
  {\frontmatter@RRAPformat{\produce@RRAP{*#1\href{mailto:#2}{#2}}}\frontmatter@RRAPformat}
  {}{}
}%
\makeatother
\begin{document}

\preprint{AIP/123-QED}

\title[]{Unsupervised Super-Resolution Data Assimilation Using Conditional Variational Autoencoders with Estimating Background Covariances via Super-Resolution}
\author{Yuki Yasuda}
 \email{yasuda.yuki@scrc.iir.isct.ac.jp}
\author{Ryo Onishi}%
 \email{onishi.ryo@scrc.iir.isct.ac.jp}
\affiliation{ 
Supercomputing Research Center, Institute of Integrated Research, Institute of Science Tokyo, \\
2-12-1 Ookayama, Meguro-ku, Tokyo 1528550, Japan.
}%

\date{\today}

\begin{abstract}
This study proposes a theory of unsupervised super-resolution data assimilation (SRDA) using conditional variational autoencoders (CVAEs). We derive an evidence lower bound for unsupervised learning, showing that our theory is an extension of a traditional data assimilation (DA) method, namely the three-dimensional variational (3D-Var) formalism. In contrast to 3D-Var, our theory exploits the non-locality of super-resolution (SR) to learn background covariances without explicitly imposing them for assimilating distant observations. For linear SR, SR operators serve as background error covariance matrices,whereas for nonlinear SR, error backpropagation through SR neural networks induces covariance structures in inference. SRDA can naturally be realized with CVAEs because the loss function for CVAEs is generally an evidence lower bound. By incorporating the SR neural network into the CVAE, the encoder estimates the high-resolution (HR) analysis from HR observations and low-resolution forecasts. The decoder acts as the observation operator by reconstructing the HR observations from the estimated HR analysis. The effectiveness of SRDA was evaluated through numerical experiments using an idealized barotropic ocean jet system. Compared to inference with an ensemble Kalman filter, SRDA demonstrated superior accuracy in HR inference. SRDA was also computationally efficient because it does not require HR numerical integration or ensemble calculations. The findings of this study provide a theoretical basis for integrating SR and DA, which will stimulate further research in this direction.
\end{abstract}

\maketitle

\section{\label{sec:introduction}Introduction}

Data assimilation (DA) is an essential technique for various modeling in computational fluid dynamics (CFD).\cite{Asch+2016SIAM, Carrassi+2018WIRE} Fluid systems generally exhibit chaotic behavior, where errors in initial conditions grow over time. Uncertainties in model parameters may also lead to discrepancies from reality. DA reduces these errors by correcting forecast values or model parameters based on observational data, thereby being essential for accurate predictions, such as numerical weather prediction.

Recent advances in deep learning are impacting various scientific fields and are also found in DA. One reason is that DA and deep learning use backpropagation based on adjoint equations.\cite{Abarbanel+2018NC, Chen+2018NIPS, Geer2021PTRS} Another reason is the use of the Bayesian approach.\cite{Carrassi+2018WIRE, Geer2021PTRS, Bocquet+2020FDS} This approach models the true states of fluid systems by probability distributions based on known data, such as observations and fluid model outputs.

A typical neural network using the Bayesian approach is the variational autoencoder (VAE).\cite{KingmaWelling2019FTML, Kingma+2014ICLR, Rezende+2014PMLR} Like standard autoencoders,\cite{Dong+2018IEEE} a VAE consists of an encoder and a decoder, both of which are implemented by neural networks. Typically, the encoder transforms the input data into a compressed representation in the latent space, while the decoder reconstructs the original data from these latent variables. A major extension of VAEs is the conditional VAE (CVAE), which is based on conditional probability modeling.\cite{Kingma+2014, Sohn+2015NIPS} This conditioning allows us to model target variables, such as the true states of systems, conditioned on known data, such as observations.

Applications of VAEs in DA can be categorized into two groups. The first group employs VAEs for dimensionality reduction.\cite{Liu+2021Geophysics,Razak+2022SPEJ,Melinc+Zaplotnik2024QJQMS} In this approach, the encoder maps fluid model states into a low-dimensional latent space. Statistical DA methods, such as Kalman filters, are then applied in this space. Due to the reduced dimensionality, the computational cost for DA is significantly lowered. This scheme is referred to as latent DA and has been extensively studied using various encoder-decoder networks.\cite{Mack+2020CMAME,Amendola2021ICCS,Peyron+2021QJRMS,Cheng+2022JCP,Liu+2022EABE,Cheng+2024CMAME} The second group employs VAEs as generators that learn the probability distributions of flow fields.\cite{Groom2021QJRMS, Yang+2021JCP, Grooms+2023QJRMS} In this approach, a fluid model state is encoded, and noise is then added in the latent space, creating various potential latent variables. Ensemble members are obtained by decoding these latent variables back into the physical space. This process facilitates fast computation of covariance matrices used in statistical DA methods. A common characteristic in both groups is that VAEs do not directly perform DA; rather, VAEs are used to make statistical DA methods more efficient. We proposes a new method, categorized into the third group, where VAEs directly perform DA.

In statistical DA methods, such as Kalman filters, the estimation of background error covariance matrices is essential for inferring the true state.\cite{Asch+2016SIAM, Carrassi+2018WIRE} Generally, observations represent a state close to reality, but they are spatially sparse. In contrast, fluid model outputs (i.e., background states) include larger errors, but they are spatially dense. DA methods correct these background states with observations in a non-local manner using covariance matrices. For instance, when an observed value exists at a grid point, this observation is used to correct surrounding (i.e., slightly distant) background values that are correlated with the background at that point. Hereafter, the term ``non-local'' means that an inferred value at a grid point depends on input values in the neighborhood (i.e., at slightly distant grid points). Conversely, ``local'' means that an inferred value is computed only from the input values at the same point.

The estimation of covariance matrices is also important for DA using autoencoders, including VAEs. In latent DA, the covariance structure needs to be estimated because the statistics of the latent space are typically unknown.\cite{Melinc+Zaplotnik2024QJQMS,Cheng+2022JCP} More generally, the background error distributions are non-Gaussian and can be estimated by VAEs.\cite{Xiao+2024arXiv} The non-Gaussian distributions learned by VAEs are then incorporated into a traditional DA method,\cite{Xiao+2024arXiv} namely, the three-dimensional variational (3D-Var)\cite{Lorenc1986QJRMS} formalism. In the present study, we focus on the non-locality of super-resolution (SR) and utilize SR operators for estimating covariance structures.

SR is originally a technique to enhance the resolution of images. In computer vision, deep learning-based SR has been actively studied.\cite{Ha+2019, Anwar+2020ACMCS, Lepcha+2023IF} A key characteristic of SR is its non-local nature:\cite{Dong+2014ECCV,Gu+Dong2021CVPR,Chen+2023CVPR,Zhou+2023ICCV} neural networks focus on input pixels at low resolution (LR) that are distant from a target pixel at high resolution (HR), but correlated to this target, generating high-quality HR images.

Recently, SR has been integrated into DA, referred to as super-resolution data assimilation (SRDA).\cite{Barthelemy+2022OD, Yasuda+Onishi2023JAMES} One reason for this integration is the increasing availability of HR observations.\cite{Imaoka+2010IEEE, Durand+2010IEEE, CIFELLI+2018JMSJ} In typical DA methods, HR observations are assimilated using HR fluid models,\cite{Li+2019JGR, Honda+2022JAMES} which require a large amount of computational resources for time integration. In contrast, in SRDA frameworks, background states are computed with LR fluid models. The LR results are then super-resolved to HR using neural networks while assimilating HR observations into these super-resolved backgrounds. SRDA methods are computationally efficient because they do not require numerical integration of HR fluid models.

One advantage of combining SR with DA is that advanced techniques developed in SR can be applied in DA. Numerous SR studies have been conducted not only in computer vision\cite{Ha+2019, Anwar+2020ACMCS, Lepcha+2023IF} but also in fluid dynamics.\cite{Fukami+2023TCFD} In contrast, DA using deep learning has only recently become active.\cite{Cheng+2023IEEE} As for SRDA, several research questions remain, such as those regarding its theoretical background. Indeed, previous studies on SRDA\cite{Barthelemy+2022OD, Yasuda+Onishi2023JAMES} did not utilize the non-locality of SR. If theoretical backgrounds for combining SR with DA are clarified, further integration of both will be promoted.

This study proposes a theory of unsupervised SRDA using CVAEs and shows that SR operators can be used to estimate background covariances. This result provides a theoretical foundation for combining SR with DA. We first review an evidence lower bound (ELBO) for DA (Section \ref{sec:preliminaries}), which serves as the loss function for unsupervised learning. We then develop a new theory of DA using the ELBO and show that this theory is an extension of the 3D-Var (Section \ref{sec:da-elbo}). Similar to the 3D-Var, our theory requires non-diagonal background error covariance matrices. We further extend our theory by incorporating SR for non-local inference, which eliminates the need to estimate non-diagonal covariance matrices (Section \ref{sec:srda-elbo}). To evaluate the proposed SRDA, numerical experiments are conducted using an idealized ocean jet system (Section \ref{sec:methods}). The results are analyzed through comparison with an ensemble Kalman filter (Section \ref{sec:results-discussion}). Conclusions are presented in Section \ref{sec:conclusions}.

\section{\label{sec:preliminaries}Review of ELBO for DA}

We reformulate an evidence lower bound (ELBO),\cite{OrmerodWand2010AS, Zhu+2014JMLR, Ghimire+2017MICCAI} which is the objective function in the proposed theory. ELBO is also known as the variational lower bound.

The present study considers assimilation at a single time step. In this context, $\bm{x}$, $\bm{y}$, and $\bm{z}$ represent a background state, observations, and the true state, respectively, where $\bm{x} \in \mathbb{R}^{n_x}$, $\bm{y} \in \mathbb{R}^{n_y}$, and $\bm{z} \in \mathbb{R}^{n_z}$. Specifically, $\bm{x}$ is a prior forecast from an underlying CFD model, $\bm{y}$ represents spatially sparse observations, and $\bm{z}$ is the true state vector in the physical space. The objective of DA is to infer $\bm{z}$ from $\bm{x}$ and $\bm{y}$.

\begin{figure}
    \includegraphics[width=7cm]{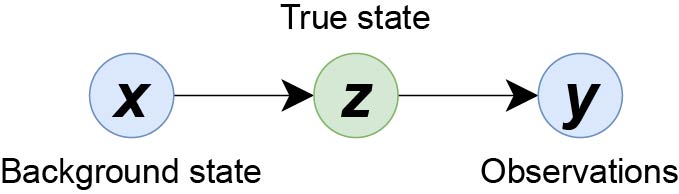}
    \caption{\label{fig:graphical-model} Probabilistic model expressed by Eq. (\ref{eq:graphical-model}). All variables are in the physical space.}
\end{figure}

Consider the probabilistic model illustrated in Fig. \ref{fig:graphical-model}, which is mathematically expressed as
\begin{equation}
    p\left(\bm{y}\mid\bm{x}\right) = \int \; p\left(\bm{y}\mid\bm{z}\right)p\left(\bm{z}\mid\bm{x}\right) \; d\bm{z}. \label{eq:graphical-model}
\end{equation}
The probability $p\left(\bm{y}\mid\bm{x}\right)$ is decomposed into two probability distributions: $p\left(\bm{y}\mid\bm{z}\right)$ describes the probability of observing $\bm{y}$ when the true state is $\bm{z}$, and $p\left(\bm{z}\mid\bm{x}\right)$ is the prior distribution for $\bm{z}$ given the model forecast $\bm{x}$.

The ELBO is derived by introducing an approximate posterior distribution $q(\bm{z}\mid \bm{x},\bm{y})$:\cite{OrmerodWand2010AS, Zhu+2014JMLR, Ghimire+2017MICCAI}
\begin{subequations}
    \label{eq:elbo-general-whole}
\begin{eqnarray}
    \ln p\left(\bm{y}\mid\bm{x}\right) &=& \ln \left[ \int \; q\left(\bm{z}\mid \bm{x},\bm{y}\right) \frac{p\left(\bm{y}\mid \bm{z}\right) p\left(\bm{z} \mid \bm{x}\right)}{q\left(\bm{z}\mid \bm{x},\bm{y}\right)} \; d\bm{z} \right], \label{eq:elbo-general-1}\\
    &\ge& \int \; q\left(\bm{z}\mid \bm{x},\bm{y}\right) \ln \left[  \frac{p\left(\bm{y}\mid \bm{z}\right) p\left(\bm{z} \mid \bm{x}\right)}{q\left(\bm{z}\mid \bm{x},\bm{y}\right)} \right] \; d\bm{z}, \label{eq:elbo-general-2}\\
    &=& \underbrace{\mathbb{E}_q \left[ \ln p\left(\bm{y}\mid \bm{z}\right) \right]}_{\text{reconstruction error}} - \underbrace{\mathbb{E}_q \left[ \ln \frac{q \left(\bm{z}\mid \bm{x},\bm{y}\right)}{p \left(\bm{z} \mid \bm{x}\right)} \right]}_{\text{KL divergence}} =: -l,\label{eq:elbo-general-3}
\end{eqnarray}
\end{subequations}
where $\mathbb{E}_q$ is the expectation operator with respect to $q \left(\bm{z}\mid \bm{x},\bm{y}\right)$. Jensen's inequality is applied in Eq. (\ref{eq:elbo-general-2}), where the equality holds only when $q(\bm{z}\mid \bm{x},\bm{y}) = p(\bm{z}\mid \bm{x},\bm{y})$, that is, when the approximate posterior is equal to the exact posterior. In Eq. (\ref{eq:elbo-general-3}), the first term is referred to as the reconstruction error for $\bm{y}$, and the second term is known as the Kullback-Leibler (KL) divergence. The final line gives the ELBO ($= -l$), which is a lower bound for the log-likelihood, $\ln p\left(\bm{y}\mid\bm{x}\right)$.

DA is usually performed via maximum likelihood estimation.\cite{Asch+2016SIAM, Carrassi+2018WIRE} Instead of maximizing the log-likelihood [i.e., $\ln p\left(\bm{y}\mid\bm{x}\right)$], we maximize the ELBO, or equivalently minimize $l$ in Eq. (\ref{eq:elbo-general-3}) to obtain $q(\bm{z}\mid \bm{x},\bm{y})$. The true state $\bm{z}$ is then inferred from $\bm{x}$ and $\bm{y}$ through $q(\bm{z}\mid \bm{x},\bm{y})$.

\section{\label{sec:da-elbo}DA using the ELBO under Gaussianity}

Starting from the ELBO, we derive a new objective function that can be used to train CVAEs in an unsupervised manner. This loss function requires non-diagonal background error covariance matrices, similar to those in 3D-Var.

\subsection{\label{subsec:transform-elbo-3dvar} Transformation of the ELBO under Gaussianity}

We make two assumptions to derive the loss function for DA: (A) the ELBO in Eq. (\ref{eq:elbo-general-3}) and (B) the following Gaussian distributions:
\begin{subequations}
    \label{eq:gaussian-whole}
\begin{eqnarray}
    q\left(\bm{z}\mid \bm{x}, \bm{y}\right) &=& \frac{1}{\sqrt{(2\pi)^{n_z} \det V(\bm{x},\bm{y})}}\; \exp \left( -\frac{1}{2} \left\lVert \bm{z} - \mathsf{A}(\bm{x}, \bm{y}) \right\rVert_{V(\bm{x},\bm{y})^{-1}}^2 \right), \label{eq:gaussian-posterior}\\
    p\left(\bm{z} \mid \bm{x}\right) &=& \frac{1}{\sqrt{(2\pi)^{n_z} \det B}} \; \exp \left( -\frac{1}{2} \left\lVert \bm{z} - \bm{x}\right\rVert_{B^{-1}}^2 \right), \label{eq:gaussian-prior}\\
    p\left(\bm{y}\mid \bm{z}\right) &=& \frac{1}{\sqrt{(2\pi)^{n_y} \det R}} \exp \left(-\frac{1}{2} \left\lVert \bm{y} - \mathsf{H}(\bm{z})\right\rVert_{R^{-1}}^2\right), \label{eq:gaussian-reconstructor}
\end{eqnarray}
\end{subequations}
where capital letters (e.g., $B$) denote matrices and sans-serif letters (e.g., $\mathsf{A}$) denote nonlinear functions that return vectors. The norm $\lVert\cdot\rVert$ represents the Mahalanobis distance: $\left\lVert \bm{z} \right\rVert_{B^{-1}}^2 = \bm{z}^{\text{T}} B^{-1} \bm{z}$ and $\left\lVert \bm{z}\right\rVert^2 = \bm{z}^{\text{T}}\bm{z}$, where an identity matrix is used when no subscript is present. In Eq. (\ref{eq:gaussian-posterior}), we assume that the mean $\mathsf{A}(\bm{x}, \bm{y})$ is the analysis state vector. The covariance matrix $V(\bm{x},\bm{y})$ describes the uncertainty of this analysis. In Eq. (\ref{eq:gaussian-prior}), $\bm{x}$ and $\bm{z}$ belong to the same vector space (i.e., $n_x = n_z$). This assumption is typical for state-space models.\cite{Bishop2006Book} In Eq. (\ref{eq:gaussian-reconstructor}), the observation operator $\mathsf{H}$  converts the true state into the observation space. In Eqs. (\ref{eq:gaussian-prior}) and (\ref{eq:gaussian-reconstructor}), $B$ and $R$ denote the prescribed background- and observation-error covariance matrices, respectively.

The ELBO is transformed by substituting Eqs. (\ref{eq:gaussian-posterior})--(\ref{eq:gaussian-reconstructor}) into Eq. (\ref{eq:elbo-general-3}). The first term, the reconstruction error in Eq. (\ref{eq:elbo-general-3}), becomes
\begin{subequations}
    \label{eq:transform-reconst-error-while}
\begin{eqnarray}
    \mathbb{E}_q \left[ \ln p\left(\bm{y}\mid \bm{z}\right) \right] &=& \mathbb{E}_q \left[ -\frac{1}{2} \left\lVert \bm{y} - \mathsf{H}(\bm{z})\right\rVert_{R^{-1}}^2 - \frac{n_y}{2} \ln (2\pi) - \frac{1}{2} \ln \left(\det R\right)\right], \label{eq:transform-reconst-error-1}\\
    &\approx& - \frac{1}{n_{\text{sample}}} \frac{1}{2} \sum_{\text{sample}} \left\lVert \bm{y} - \mathsf{H}(\bm{\hat{z}})\right\rVert_{R^{-1}}^2 + \text{const}, \label{eq:transform-reconst-error-2}
\end{eqnarray}
\end{subequations}
where
\begin{equation}
    \bm{\hat{z}} = \mathsf{A}(\bm{x}, \bm{y}) + \bm{\varepsilon}^{\text{T}} V(\bm{x},\bm{y})^{1/2} = \bm{a} + \bm{\varepsilon}^{\text{T}} V^{1/2}. \label{eq:def-hat-z}
\end{equation}
Here, $\bm{a}$ and $V$ denote the vector and matrix given by $\mathsf{A}(\bm{x}, \bm{y})$ and $V(\bm{x},\bm{y})$, respectively, $\bm{\varepsilon}$ is an $n_z$-dimensional standard normal variable, and $V^{1/2}$ is obtained by Cholesky decomposition. In Eq. (\ref{eq:transform-reconst-error-2}), the expected value is approximated by the sample mean using a Monte Carlo method (i.e., the reparameterization trick),\cite{Kingma+2014ICLR} where each realization $\bm{\hat{z}}$ is sampled from $q\left(\bm{z}\mid \bm{x}, \bm{y}\right)$ as in Eq. (\ref{eq:def-hat-z}). In the actual implementation, the sample size $n_{\text{sample}}$ can be set to 1 if the batch size is sufficiently large.\cite{Kingma+2014ICLR} Equation (\ref{eq:transform-reconst-error-2}) indicates that the reconstruction error measures the difference between $\bm{y}$ and $\mathsf{H}(\bm{\hat{z}})$ in the observation space.

The KL divergence, the second term in Eq. (\ref{eq:elbo-general-3}), is transformed as follows:
\begin{subequations}
    \label{eq:transform-kl-div-whole}
\begin{eqnarray}
    \mathbb{E}_q \left[ \ln \frac{q \left(\bm{z}\mid \bm{x},\bm{y}\right)}{p \left(\bm{z} \mid \bm{x}\right)} \right] &=& \int d\bm{z} \; q \left(\bm{z}\mid \bm{x},\bm{y}\right) \left\{ -\frac{1}{2} \ln \left[\det V(\bm{x},\bm{y})\right] -\frac{1}{2} \left\lVert \bm{z} - \mathsf{A}(\bm{x}, \bm{y})\right\rVert_{V(\bm{x},\bm{y})^{-1}}^2 \right\}, \notag \\
      &\quad&\quad - \int d\bm{z} \; q \left(\bm{z}\mid \bm{x},\bm{y}\right) \left\{ -\frac{1}{2} \ln \left(\det B\right)  -\frac{1}{2} \left\lVert \bm{z} - \bm{x} \right\rVert_{B^{-1}}^2 \right\}, \label{eq:transform-kl-div-1} \\
      &=& \frac{1}{\sqrt{(2\pi)^{n_z}}} \int d\widetilde{\bm{z}} \; \exp \left(-\frac{\lVert \widetilde{\bm{z}}\rVert^2}{2}\right) \Bigg\{ - \frac{1}{2} \ln \left[\frac{\det V(\bm{x},\bm{y})}{\det B}\right] \notag \\
      &\quad&\quad + \frac{1}{2} \left\lVert \left[ V(\bm{x},\bm{y})^{1/2} \widetilde{\bm{z}} + \mathsf{A}(\bm{x}, \bm{y}) \right] - \bm{x} \right\rVert_{B^{-1}}^2 \Bigg\}, \label{eq:transform-kl-div-2} \\
      &=& \frac{1}{2} \left\lVert \mathsf{A}(\bm{x}, \bm{y}) - \bm{x} \right\rVert_{B^{-1}}^2 + \frac{1}{2} \left\{ \operatorname{tr}\left[B^{-1}V(\bm{x},\bm{y}) \right] - \ln \left[\frac{\det V(\bm{x},\bm{y})}{\det B}\right]\right\}. \label{eq:transform-kl-div-3}
\end{eqnarray}
\end{subequations}
We introduce an auxiliary variable $\widetilde{\bm{z}} = V(\bm{x},\bm{y})^{-1/2} \left[\bm{z} - \mathsf{A}(\bm{x}, \bm{y})\right]$ to perform the integral. For brevity, constant terms are omitted in Eqs. (\ref{eq:transform-kl-div-2}) and (\ref{eq:transform-kl-div-3}). The first term in Eq. (\ref{eq:transform-kl-div-3}) represents the difference between the analysis $\mathsf{A}(\bm{x}, \bm{y})$ and the background state $\bm{x}$, while the second term is a convex function of $V(\bm{x},\bm{y})$ that takes the minimum when $V(\bm{x},\bm{y})=B$.

By substituting Eqs. (\ref{eq:transform-reconst-error-2}) and (\ref{eq:transform-kl-div-3}) into Eq. (\ref{eq:elbo-general-3}), we obtain the ELBO under Gaussianity:
\begin{eqnarray}
    l_{\text{Gauss}} &=& \frac{1}{2} \left\lVert \bm{y} - \mathsf{H}(\bm{\hat{z}})\right\rVert_{R^{-1}}^2 + \frac{1}{2} \left\lVert \mathsf{A}(\bm{x}, \bm{y}) - \bm{x} \right\rVert_{B^{-1}}^2 +\frac{1}{2} \left\{ \operatorname{tr}\left[B^{-1}V(\bm{x}, \bm{y}) \right] - \ln \left[\frac{\det V(\bm{x}, \bm{y})}{\det B}\right]\right\}, \label{eq:elbo-gauss}
\end{eqnarray}
where constant terms are omitted and the sample size $n_{\text{sample}}$ in Eq. (\ref{eq:transform-reconst-error-2}) is set to 1.

\subsection{\label{subsec:reduction-to-3dvar} Reduction of the ELBO to the 3D-Var objective function}

We show that $l_{\text{Gauss}}$ can be reduced to the 3D-Var objective function. This suggests that DA can be performed by minimizing $l_{\text{Gauss}}$.

To show this fact, we make further one assumption: the covariance $V(\bm{x}, \bm{y})$ is a constant matrix, where each component is quite small. The variable $\bm{\hat{z}}$ in Eq. (\ref{eq:def-hat-z}) is then approximated as $\bm{\hat{z}} = \mathsf{A}(\bm{x}, \bm{y}) = \bm{a}$, and the third term in Eq. (\ref{eq:elbo-gauss}) becomes constant. Consequently, $l_{\text{Gauss}}$ is reduced to the 3D-Var objective function $l_{\text{3D-Var}}$:
\begin{equation}
    l_{\text{3D-Var}} = \frac{1}{2}\left\lVert \bm{y} - \mathsf{H}(\bm{a})\right\rVert_{R^{-1}}^2 + \frac{1}{2}\left\lVert\bm{a} - \bm{x} \right\rVert_{B^{-1}}^2, \label{eq:elbo-3d-var}
\end{equation}
where $\bm{a}$ denotes the analysis vector given by the function $\mathsf{A}$ from $\bm{x}$ and $\bm{y}$ [i.e., $\bm{a}=\mathsf{A}(\bm{x},\bm{y})$]. This $\bm{a}$ is obtained by minimizing $l_{\text{3D-Var}}$. The first term of $l_{\text{3D-Var}}$, which comes from the reconstruction error in Eq. (\ref{eq:elbo-general-3}), represents the difference between the observation vector $\bm{y}$ and the analysis $\bm{a}$ via the observation operator $\mathsf{H}$. The second term of $l_{\text{3D-Var}}$, which comes from the KL divergence in Eq. (\ref{eq:elbo-general-3}), represents the difference between the analysis $\bm{a}$ and the background state $\bm{x}$. By minimizing $l_{\text{3D-Var}}$, $\bm{x}$ and $\bm{y}$ are integrated to estimate the true state as $\bm{a}$. Since $l_{\text{3D-Var}}$ is derived from the ELBO [i.e., $l_{\text{Gauss}}$ in Eq. (\ref{eq:elbo-gauss})], the ELBO describes the balance between $\bm{x}$ and $\bm{y}$ in estimating the true state.

The minimizer of $l_{\text{3D-Var}}$ is easily derived when the observation operator is a matrix $H$:
\begin{subequations}
    \label{eq:minimizer-3d-var-whole}
\begin{eqnarray}
    \bm{a} &=& \bm{x} + K \left(\bm{y} - H\bm{x}\right), \label{eq:minimizer-3d-var}\\
    K &=& BH^{\text{T}}\left(R + H B H^{\text{T}} \right)^{-1}, \label{eq:kalman-gain}
\end{eqnarray}
\end{subequations}
where $K$ is known as the Kalman gain matrix.\cite{Asch+2016SIAM} These equations describe the non-locality of DA: an observation at one grid point is incorporated into the surrounding inferred values. This non-locality typically originates from the non-zero off-diagonal elements of the covariance matrix $B$. If $B$ is diagonal (i.e., all off-diagonal elements are zero), DA becomes local inference, and spatially sparse observations are incorporated into the inference only at their located grid points.

\subsection{\label{subsec:3dvar-implementation-cvae} Implementation of DA using CVAE}

DA can be performed by minimizing $l_{\text{Gauss}}$ in Eq. (\ref{eq:elbo-gauss}). However, this minimization at each assimilation time step would be computationally expensive. We propose an efficient approach to this minimization using a trained CVAE.

The loss function for VAEs, including CVAEs, is generally an ELBO.\cite{KingmaWelling2019FTML, Kingma+2014ICLR, Rezende+2014PMLR, Kingma+2014, Sohn+2015NIPS} This can be understood from the ELBO formula [Eq. (\ref{eq:elbo-gauss})] and naturally leads to the use of CVAEs rather than other models (e.g., generative adversarial networks\cite{Creswell+2018IEEE}) that do not have ELBOs as their loss functions. To evaluate $l_{\text{Gauss}}$ in Eq. (\ref{eq:elbo-gauss}), the analysis $\bm{a}$ and its covariance $V$ are first computed from given $\bm{x}$ and $\bm{y}$; the observation vector $\bm{\hat{y}}$ [$=\mathsf{H}(\bm{\hat{z}})$] is then reconstructed from $\bm{a}$ and $V$ via Eq. (\ref{eq:def-hat-z}). The value of $l_{\text{Gauss}}$ decreases as $\bm{\hat{y}}$ becomes closer to the input $\bm{y}$. The architecture that reconstructs the input is a characteristic of autoencoders.\cite{Dong+2018IEEE} The second and third terms in Eq. (\ref{eq:elbo-gauss}) are the conditions for the estimated true state $\bm{\hat{z}}$. To incorporate these conditions, we use a CVAE, which is trained by minimizing the loss function $l_{\text{Gauss}}$. This training is unsupervised because $l_{\text{Gauss}}$ does not require the ground truth for $\bm{z}$. This unsupervised learning is achieved through the use of the ELBO because $l_{\text{Gauss}}$ is derived from the ELBO in Eq. (\ref{eq:elbo-general-3}).

\begin{figure}
    \includegraphics[width=12cm]{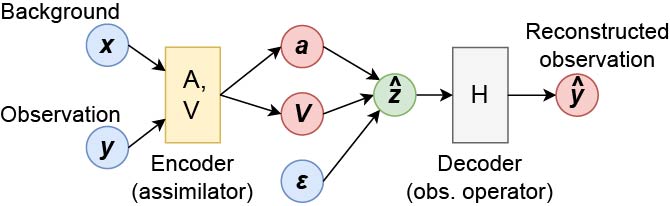}
    \caption{\label{fig:schematic_DA} Architecture of the CVAE for DA. The encoder estimates the analysis vector $\bm{a}$ and its covariance $V$, both of which give a sample of the true state $\bm{\hat{z}}$ in Eq. (\ref{eq:def-hat-z}). The decoder acts as the observation operator and reconstructs the observation vector $\bm{\hat{y}} = \mathsf{H}(\bm{\hat{z}})$.}
\end{figure}

The CVAE consists of an encoder and a decoder (Fig. \ref{fig:schematic_DA}). The encoder estimates $\bm{a}$ and $V$ from the input $\bm{x}$ and $\bm{y}$, while the decoder reconstructs $\bm{\hat{y}}$. In other words, the encoder performs assimilation, while the decoder acts as the observation operator. The encoder learns how to integrate $\bm{x}$ and $\bm{y}$, yielding the analysis $\bm{a}$, while the decoder learns how to convert $\bm{a}$ into the observation space. Once the CVAE is trained, the encoder performs DA without minimizing $l_{\text{Gauss}}$ as long as the statistical properties of $\bm{x}$ and $\bm{y}$ do not change significantly.

The covariance matrices $R$ and $B$ in $l_{\text{Gauss}}$ are treated as prescribed parameters (i.e., hyperparameters). Since it is difficult to determine many hyperparameters, these matrices are usually assumed to be proportional to identity matrices (i.e., having zero off-diagonal elements).\cite{KingmaWelling2019FTML, Kingma+2014ICLR, Rezende+2014PMLR, Kingma+2014, Sohn+2015NIPS} However, the off-diagonal elements of $B$ are essential for the non-local inference in DA (Section \ref{subsec:reduction-to-3dvar}). We adopt super-resolution (SR) to realize non-local inference using CVAEs, leading to a new theory of SRDA.

\section{\label{sec:srda-elbo} SRDA using the ELBO}

\subsection{\label{subsec:integration-sr-into-da} Integration of SR into DA}

To develop a theory of SRDA, we first specify the resolutions of a background state $\bm{x}$, observations $\bm{y}$, and the true state $\bm{z}$: $\bm{x}$ is defined at LR grid points, while $\bm{y}$ and $\bm{z}$ are defined at HR grid points. The vector forms are obtained by arranging three-dimensional grid point values into columns. Observations are not available at all locations due to various factors, such as measurement errors. These missing grid points are excluded when arranging observations into the column vector $\bm{y}$.

We make two assumptions to derive the loss function for SRDA: (A) the ELBO in Eq. (\ref{eq:elbo-general-3}) and (B') the following Gaussian distributions. The first assumption is the same as in the theory for DA in Section \ref{sec:da-elbo}, while the second assumption is a modification from Eqs. (\ref{eq:gaussian-posterior})--(\ref{eq:gaussian-reconstructor}):
\begin{subequations}
    \label{eq:gaussian-whole-srda}
\begin{eqnarray}
    q\left(\bm{z}\mid \bm{x}, \bm{y}\right) &=& \frac{1}{\sqrt{(2\pi)^{n_z} \det V(\bm{x}, \bm{y})}} \;\exp \Bigg\{ -\frac{1}{2} \left\lVert\bm{z} - \mathsf{A}_{\text{HR}}(\bm{x}, \bm{y})\right\rVert^2_{V(\bm{x}, \bm{y})^{-1}}\Bigg\}, \label{eq:gaussian-posterior-srda}\\
    p\left(\bm{z} \mid \bm{x}\right) &=& \frac{1}{(\sqrt{2\pi b})^{n_z}} \;\exp \left\{-\frac{1}{2b} \left\lVert \bm{z} - \mathsf{F}(\bm{x})\right\rVert^2\right\}, \label{eq:gaussian-prior-srda}\\
    p\left(\bm{y}\mid \bm{z}\right) &=& \frac{1}{(\sqrt{2\pi r})^{n_y}} \;\exp \left\{-\frac{1}{2r} \left\lVert \bm{y} - \mathsf{H}(\bm{z})\right\rVert^2\right\}, \label{eq:gaussian-reconstructor-srda}
\end{eqnarray}
\end{subequations}
where
\begin{equation}
    \mathsf{A}_{\text{HR}} = \mathsf{F} \circ \mathsf{A}_{\text{LR}}. \label{eq:def-operator-A-hr}
\end{equation}
We first explain the differences between Eqs. (\ref{eq:gaussian-whole}) and (\ref{eq:gaussian-whole-srda}) due to $B$ and $R$ being diagonal: $B = b I_{n_z}$ and $R = r I_{n_y}$, where $b$ and $r$ are prescribed positive parameters and $I_n$ is an identity matrix of size $n$. The exponents are then expressed as $\left\lVert \bm{z} - \mathsf{F}(\bm{x})\right\rVert_{B^{-1}}^2 = \left\lVert \bm{z} - \mathsf{F}(\bm{x})\right\rVert^2/b$ and $\left\lVert \bm{y} - \mathsf{H}(\bm{z})\right\rVert_{R^{-1}}^2 = \left\lVert \bm{y} - \mathsf{H}(\bm{z})\right\rVert^2 / r$ in Eqs. (\ref{eq:gaussian-prior-srda}) and (\ref{eq:gaussian-reconstructor-srda}), respectively. Note that the norm $\lVert \cdot \rVert$ with no subscript means the use of identity matrices, such as $\lVert \bm{z}\rVert^2 = \bm{z}^{\text{T}} I_{n_z} \bm{z} = \bm{z}^{\text{T}}\bm{z}$. The determinants are expressed as $\det B = b^{n_z}$ and $\det R = r^{n_y}$. We also take the covariance $V(\bm{x}, \bm{y})$ in Eq. (\ref{eq:gaussian-posterior-srda}) to be diagonal:
\begin{equation}
    V(\bm{x}, \bm{y}) = \text{diag}\left[v_1(\bm{x}, \bm{y}), \ldots, v_{n_z}(\bm{x}, \bm{y}) \right], \label{eq:def-variance-srda}
\end{equation}
where each variance $v_i(\bm{x}, \bm{y})$ ($i=1,\ldots,n_z$) is positive and determined by $\bm{x}$ and $\bm{y}$.

There are two essential changes from Eqs. (\ref{eq:gaussian-whole}) to (\ref{eq:gaussian-whole-srda}). First, an SR operator $\mathsf{F}$ is introduced in Eq. (\ref{eq:gaussian-prior-srda}), mapping an $n_x$-dimensional vector into an $n_z$-dimensional vector. Since $\bm{x}$ and $\bm{z}$ are defined at LR and HR grid points, respectively (i.e., $n_x < n_z$), this mapping performs SR. Comparing Eqs. (\ref{eq:gaussian-prior}) and (\ref{eq:gaussian-prior-srda}), the mean of the prior distribution $p\left(\bm{z} \mid \bm{x}\right)$ is replaced with the super-resolved background state $\mathsf{F}(\bm{x})$. Second, the analysis state is also super-resolved using $\mathsf{F}$. The function $\mathsf{A}$ in Eq. (\ref{eq:gaussian-posterior}) is replaced with a composition $\mathsf{A}_{\text{HR}} = \mathsf{F} \circ \mathsf{A}_{\text{LR}}$ in Eq. (\ref{eq:gaussian-posterior-srda}): the LR analysis is first calculated and then super-resolved.

The ELBO for SRDA is obtained by substituting Eqs. (\ref{eq:gaussian-posterior-srda})--(\ref{eq:gaussian-reconstructor-srda}) into Eq. (\ref{eq:elbo-general-3}), as in Section \ref{subsec:transform-elbo-3dvar}:
\begin{eqnarray}    
    l_{\text{SRDA}} &=& \frac{1}{2r} \left\lVert \bm{y} - \mathsf{H}(\bm{\hat{z}})\right\rVert^2 + \frac{1}{2b} \left\lVert \mathsf{A}_{\text{HR}}(\bm{x},\bm{y}) - \mathsf{F}(\bm{x})\right\rVert^2 + \frac{1}{2} \sum_{i=1}^{n_z}\left\{\frac{v_i(\bm{x},\bm{y})}{b} - \ln\left[\frac{v_i(\bm{x},\bm{y})}{b}\right] \right\}, \label{eq:elbo-srda}
\end{eqnarray}
where
\begin{subequations}
\begin{eqnarray}
    \bm{\hat{z}} &=& \mathsf{A}_{\text{HR}}(\bm{x}, \bm{y}) + \bm{\varepsilon}^{\text{T}} V(\bm{x},\bm{y})^{1/2} = \bm{a}_{\text{HR}} + \bm{\varepsilon}^{\text{T}} V^{1/2}, \label{eq:def-hat-z-srda} \\
    \bm{a}_{\text{HR}} &=& \mathsf{A}_{\text{HR}}(\bm{x},\bm{y}) = \mathsf{F}(\bm{a}_{\text{LR}}), \label{eq:def-a-hr}\\
    \bm{a}_{\text{LR}} &=& \mathsf{A}_{\text{LR}}(\bm{x},\bm{y}), \label{eq:def-a-lr}
\end{eqnarray}
\end{subequations}
Eqs. (\ref{eq:elbo-srda}) and (\ref{eq:def-hat-z-srda}) correspond to Eqs. (\ref{eq:elbo-gauss}) and (\ref{eq:def-hat-z}), respectively, where $V$ in Eq. (\ref{eq:def-hat-z-srda}) is defined by Eq. (\ref{eq:def-variance-srda}). As in Eq. (\ref{eq:elbo-gauss}), the Monte Carlo method (i.e., the reparameterization trick)\cite{Kingma+2014ICLR} is applied in Eq. (\ref{eq:elbo-srda}) with the sample size being 1. In the right-hand side of Eq. (\ref{eq:def-hat-z-srda}), the arguments of $\bm{x}$ and $\bm{y}$ are omitted, and the definition of $\bm{a}_{\text{HR}}$ [Eq. (\ref{eq:def-a-hr})] is used. This $\bm{a}_{\text{HR}}$ is obtained by super-resolving $\bm{a}_{\text{LR}}$. The first term of $l_{\text{SRDA}}$ represents the difference between the observation vector $\bm{y}$ and the estimated true state $\mathsf{H}(\bm{\hat{z}})$ in the observation space. The second term is the difference between the super-resolved analysis $\bm{a}_{\text{HR}}$ [$=\mathsf{A}_{\text{HR}}(\bm{x}, \bm{y})$] and the super-resolved background $\mathsf{F}(\bm{x})$. The third term is a convex function of $v_i$ that takes the minimum when $v_i(\bm{x},\bm{y}) = b$. The ELBO (i.e., $l_{\text{SRDA}}$) describes the balance between the observations and the super-resolved background state in estimating the HR true state.

\subsection{\label{subsec:srda-linear-sr} Case of linear SR}

The loss function in Eq. (\ref{eq:elbo-srda}) is applicable to both linear and nonlinear SR. This section and the next (Section \ref{subsec:srda-nonlinear-sr}) are for interpreting this loss function through examining the effects of non-locality of SR in the linear and nonlinear regimes, respectively. Here, we show that linear SR operators act as background error covariance matrices.

Linear SR operators were frequently used in SR studies, especially in theoretical studies, before the emergence of deep learning.\cite{Park+2003IEEE} When generation processes of LR images from HR images are linear (e.g., subsampling or blurring), the inverse processes, namely the SR processes, are also linear.\cite{Park+2003IEEE} An SR operator is denoted by a matrix $F$ of size $n_z \times n_x$ ($n_z > n_x$) that maps $n_x$-dimensional LR vectors into $n_z$-dimensional HR vectors.

To simplify the discussion, we take the variances $v_i$ ($i=1, \ldots, n_z$) to be constant and small values, as in Section \ref{subsec:reduction-to-3dvar}. The estimated true state $\bm{\hat{z}}$ in Eq. (\ref{eq:def-hat-z-srda}) is then approximated as $\bm{\hat{z}} = F \bm{a}_{\text{LR}}$. Further, the third term in Eq. (\ref{eq:elbo-srda}) is ignored because it becomes constant.

\begin{figure}
    \includegraphics[width=10cm]{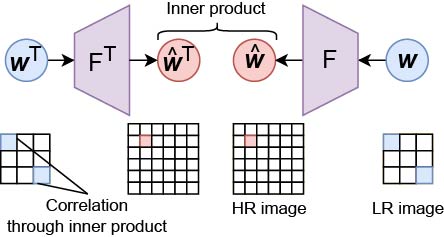}
    \caption{\label{fig:schematic_linear_SRDA} Schematic of spatial correlation through a linear SR operator $F$. This $F$ super-resolves an LR image $\bm{w}$ to the HR image $\bm{\hat{w}}$ ($=F \bm{w}$). Here, $\bm{w}$ corresponds to $\bm{a}_{\text{LR}} - \bm{x}$ in Eq. (\ref{eq:elbo-linear-srda}). The two blue pixels in $\bm{w}$ contribute to the red pixel in $\bm{\hat{w}}$. By taking an inner product $\bm{\hat{w}}^{\text{T}}\bm{\cdot}\bm{\hat{w}}$, the two blue pixels are correlated with each other.}
\end{figure}

Consequently, $l_{\text{SRDA}}$ is reduced to
\begin{equation}
    l_{\text{SRDA-L}} = \frac{1}{2r} \left\lVert \bm{y} - \mathsf{H}(F \bm{a}_{\text{LR}})\right\rVert^2 + \frac{1}{2b} \left(\bm{a}_{\text{LR}} - \bm{x}\right)^{\text{T}} \left(F^{\text{T}} F\right) \left(\bm{a}_{\text{LR}} - \bm{x}\right), \label{eq:elbo-linear-srda}
\end{equation}
where SRDA-L stands for SRDA with linear SR. The matrix $F^{\text{T}} F$ is an $n_x \times n_x$ positive semi-definite matrix. Compared to the ELBO for 3D-Var in Eq. (\ref{eq:elbo-3d-var}), $F^{\text{T}} F$ acts as the covariance matrix $B$. Due to the non-locality of SR, $F^{\text{T}} F$ is non-diagonal and describes the spatial correlations among the LR components (Fig. \ref{fig:schematic_linear_SRDA}).

\subsection{\label{subsec:srda-nonlinear-sr} Case of nonlinear SR}

In practice, SR operators are implemented as neural networks and are thus nonlinear. Since nonlinear SR encompasses linear SR, non-local inference is also achieved with nonlinear SR. In this section, non-locality is examined from a perspective of backpropagation through an SR neural network.

We consider the nonlinear version of Eq. (\ref{eq:elbo-linear-srda}):
\begin{equation}
    l_{\text{SRDA-NL}} = \frac{1}{2r} \left\lVert H\mathsf{F}(\bm{a}) - \bm{y}\right\rVert^2 + \frac{1}{2b} \left\lVert \mathsf{F}(\bm{a}) - \mathsf{F}(\bm{x})\right\rVert^2, \label{eq:elbo-nonlinear-srda}
\end{equation}
where SRDA-NL stands for SRDA with nonlinear SR. For simplicity, we set the observation operator as a matrix ($\mathsf{H}=H$) and denote $\bm{a}_{\text{LR}}$ as $\bm{a}$. The minimizer $\bm{a}$ can be estimated by gradient descent in an iterative manner:
\begin{subequations}
\begin{eqnarray}
    a^{(n+1)}_i &=& a^{(n)}_i - \eta^{(n)} \frac{\partial}{\partial a^{(n)}_i} l_{\text{SRDA-NL}}, \label{eq:gradient-descent1}\\
    \frac{\partial}{\partial a^{(n)}_i} l_{\text{SRDA-NL}} &=& \frac{1}{r} \sum_{j,k,l} \left[ H_{jk}\mathsf{F}_k(\bm{a}^{(n)}) - y_j \right] H_{jl} \frac{\partial \mathsf{F}_l(\bm{a}^{(n)})}{\partial a^{(n)}_i} \notag \\
    &+& \frac{1}{b} \sum_{j} \left[ \mathsf{F}_j(\bm{a}^{(n)}) - \mathsf{F}_j(\bm{x}) \right] \frac{\partial \mathsf{F}_j(\bm{a}^{(n)})}{\partial a^{(n)}_i}, \label{eq:gradient-descent2}
\end{eqnarray}
\end{subequations}
where a superscript, such as $(n)$, denotes the number of iterations, and $\eta^{(n)}$ is a small positive value for the $n$-th iteration.

\begin{figure}
    \centering
    \includegraphics[width=9cm]{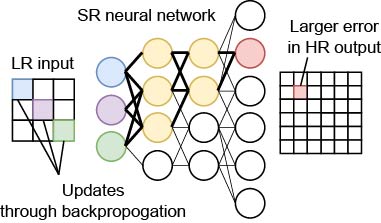}
    \caption{\label{fig:schematic_nonlinear_SRDA} Schematic of the non-local update by backpropagation through an SR neural network (i.e., a non-linear SR model). A large error (red), $H\mathsf{F}(\bm{a}^{(n)}) - \bm{y}$, propagates backward through yellow nodes in the hidden layers and updates the LR input $\bm{a}^{(n)}$ to make $H\mathsf{F}(\bm{a}^{(n)})$ closer to $\bm{y}$. See text for details.}
\end{figure}

Eqs. (\ref{eq:gradient-descent1}) and (\ref{eq:gradient-descent2}) are similar to backpropagation formulas that use mean squared errors as a loss function.\cite{Goodfellow+16textbook} The differences from the observations (i.e., errors), $H\mathsf{F}(\bm{a}^{(n)}) - \bm{y}$, propagate backward through the SR neural network $\mathsf{F}$. This backpropagation updates the current solution $\bm{a}^{(n)}$ to make $H\mathsf{F}(\bm{a}^{(n)})$ closer to $\bm{y}$ (Fig. \ref{fig:schematic_nonlinear_SRDA}). Note that all parameters of $\mathsf{F}$ remain fixed because $\mathsf{F}$ is already trained. Since trained SR neural networks generally exhibit non-locality,\cite{Dong+2014ECCV,Gu+Dong2021CVPR,Chen+2023CVPR,Zhou+2023ICCV} the backpropagation updates the analysis vector $\bm{a}$ using observations from distant grid points (Fig. \ref{fig:schematic_nonlinear_SRDA}). We demonstrate that a trained SR model actually exhibits this non-locality in Section \ref{subsec:accuracy-sr-model}. Equation (\ref{eq:gradient-descent2}) indicates that the differences from the super-resolved background state, $\mathsf{F}(\bm{a}^{(n)}) - \mathsf{F}(\bm{x})$, also induce updates through backpropagation. The ELBO [Eq. (\ref{eq:elbo-nonlinear-srda})] determines the balance between the updates from observation and background values.

Minimizing Eq. (\ref{eq:elbo-nonlinear-srda}) through backpropagation may be computationally expensive, although automatic differentiation in deep learning frameworks facilitates the calculation of $\partial \mathsf{F} / \partial \bm{a}^{(n)}$ in Eq. (\ref{eq:gradient-descent2}). As in Section \ref{subsec:3dvar-implementation-cvae}, we propose to solve the minimization problem using CVAEs.

\subsection{\label{subsec:loss-for-sr-models} Loss function for SR models}

So far, we have used the trained SR model $\mathsf{F}$. This section introduces a loss function for training $\mathsf{F}$. We do not know the true state of the system and only know the LR background $\bm{x}$ and the HR observations $\bm{y}$, where $\bm{x}$ is spatially complete and $\bm{y}$ is spatially incomplete. The SR model is trained in this setup.

Assuming that the observation operator is a projection matrix, we propose the following loss function to train SR models:
\begin{equation}
    l_{\text{SR}} = \sum_{i=1}^{n_y} \left| y_i - [P\mathsf{F}(\bm{x})]_i \right|, \label{eq:loss-sr}
\end{equation}
where $|\cdot|$ denotes the absolute value, $i$ specifies vector components, and $P$ is the projection matrix into the observation space ($\mathsf{H} = P$). Typically, $P$ consists of interpolation and/or subsampling operations. The right-hand side shows that $l_{\text{SR}}$ is an L1 norm, which is frequently used in SR studies.\cite{Ha+2019, Anwar+2020ACMCS, Lepcha+2023IF, Fukami+2023TCFD} Backpropagation is feasible because $l_{\text{SR}}$ is differentiable with respect to $\mathsf{F}(\bm{x})$. The SR model $\mathsf{F}$ is trained using $\bm{x}$ as input and $\bm{y}$ as target. When the training data size is sufficiently large, the trained SR model infers the median of $\bm{y}$ conditioned on $\bm{x}$.\cite{Hastie+2009Springer} This suggests that the SR model becomes unbiased if the median of the observation error distribution is zero.

\subsection{\label{subsec:srda-implementation-cvae} Implementation of SRDA using CVAE}

An efficient minimization of $l_{\text{SRDA}}$ in Eq. (\ref{eq:elbo-srda}) is possible with CVAEs, as in Section \ref{subsec:3dvar-implementation-cvae}, as long as the statistical properties of $\bm{x}$ and $\bm{y}$ do not change significantly.

\begin{figure}
    \includegraphics[width=14.5cm]{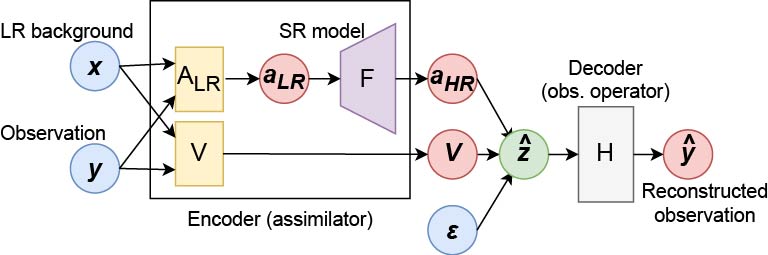}
    \caption{\label{fig:schematic_SRDA} Architecture of the CVAE for SRDA. The encoder estimates the HR analysis vector $\bm{a}_{\text{HR}}$ using the SR model. The encoder also estimates the variance $V$ for $\bm{a}_{\text{HR}}$. The variables $\bm{a}_{\text{HR}}$ and $V$ give a sample of the true state $\bm{\hat{z}}$ in Eq. (\ref{eq:def-hat-z-srda}). The decoder acts as the observation operator and reconstructs the observation vector $\bm{\hat{y}} = \mathsf{H}(\bm{\hat{z}})$.}
\end{figure}

The CVAE for SRDA consists of an encoder and a decoder (Fig. \ref{fig:schematic_SRDA}). The difference from the CVAE in Fig. \ref{fig:schematic_DA} is that the encoder incorporates an SR model and performs SR and DA simultaneously. We use the SR model in Section \ref{subsec:loss-for-sr-models}, which is trained with the loss function in Eq. (\ref{eq:loss-sr}). The encoder integrates the LR background $\bm{x}$ and the HR observations $\bm{y}$,  estimating the HR analysis $\bm{a}_{\text{HR}}$ through computing the LR analysis $\bm{a}_{\text{LR}}$. This computation of $\bm{a}_{\text{LR}}$ [$=A_{\text{LR}}(\bm{x}, \bm{y})$] emulates the result of iterative backpropagation through $\mathsf{F}$ (i.e., the result of minimizing $l_{\text{SRDA-NL}}$ in Section \ref{subsec:srda-nonlinear-sr}). The encoder also estimates the uncertainty of $\bm{a}_{\text{HR}}$ [$=\mathsf{F}(\bm{a}_{\text{LR}})$] by calculating the variance $V$ in Eq. (\ref{eq:def-variance-srda}). The decoder acts as the observation operator, as in Fig. \ref{fig:schematic_DA}. The CVAE is trained in an unsupervised manner using $l_{\text{SRDA}}$ in Eq. (\ref{eq:elbo-srda}) as the loss function.

Rigorously speaking, a different SR model should be used within the encoder because the SR model in Section \ref{subsec:loss-for-sr-models} is trained to super-resolve the LR background $\bm{x}$, which differs from the LR analysis $\bm{a}_{\text{LR}}$. This difference in input may introduce bias in $\bm{a}_{\text{LR}}$. In contrast, the output $\bm{a}_{\text{HR}}$ [$=\mathsf{F}(\bm{a}_{\text{LR}})$] is expected to be less biased because the SR model is unbiased if the observation error distribution has a zero median (Section \ref{subsec:loss-for-sr-models}). Thus, $\bm{a}_{\text{HR}}$ is used as initial conditions for LR fluid models by resizing $\bm{a}_{\text{HR}}$ to the LR (Section \ref{subsec:sequentiaSRDA-L}). One solution for making $\bm{a}_{\text{LR}}$ less biased is to conduct fine-tuning of the SR model $\mathsf{F}$ during the training of the CVAE. For simplicity, this study does not use such an advanced learning method, which remains a future research topic; instead, the parameters of $\mathsf{F}$ are fixed during the CVAE training.

\subsection{\label{subsec:sequentiaSRDA-L} Sequential SRDA using the CVAE}

We introduce a sequential SRDA method applicable to CFD simulations. The background state $\bm{x}$ is hereafter referred to as the forecast state because it is a prior forecast given by an LR fluid model. The forecast and assimilation are executed alternately, as in other statistical DA methods.\cite{Asch+2016SIAM, Carrassi+2018WIRE} This assimilation is performed using a trained CVAE's encoder.

\begin{figure}
    \centering
    \includegraphics[width=12cm]{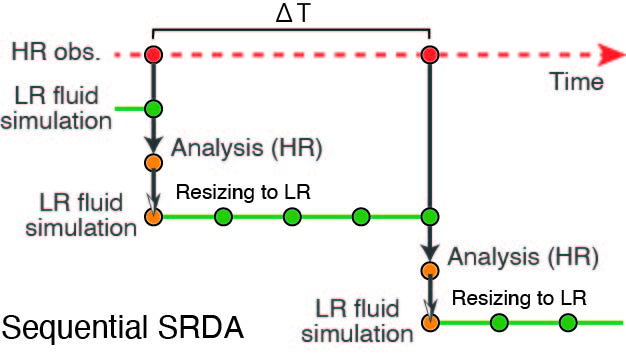}
    \caption{\label{fig:sequential_SRDA} Schematic of the sequential SRDA. ``HR obs'' denotes HR observations. A constant assimilation interval $\Delta T$ is assumed, where the output time step is a quarter of $\Delta T$ as an example.}
\end{figure}

Figure \ref{fig:sequential_SRDA} shows a schematic of the sequential SRDA. We assume a constant assimilation interval $\Delta T$. At each assimilation time step, the encoder assimilates the HR observations $\bm{y}$ into the LR forecast state $\bm{x}$, thereby estimating the HR analysis $\bm{a}_{\text{HR}}$. This $\bm{a}_{\text{HR}}$ is resized to the LR using an algebraic method, such as linear interpolation. In the next assimilation cycle, the resized analysis is used as the initial condition for the LR fluid model, and forecasts are computed by numerical integration over $\Delta T$. The forecast at the final time step is fed to the encoder to infer the subsequent HR analysis.

The decoder is not used in the sequential SRDA; rather, it is used only in the training phase to learn the observation operator. In contrast, the trained SR model can be used in the test phase (i.e., the operational phase) to super-resolve LR forecast states. This SR operation is independent of assimilation, enabling the inference of HR forecasts at any time step.

\subsection{\label{subsec:related-studies} Related studies}

We first compare the present SRDA with previous SRDA methods.\cite{Barthelemy+2022OD, Yasuda+Onishi2023JAMES} The major differences are that the present SRDA is based on unsupervised learning and makes full use of the non-locality of SR. Barth\'{e}l\'{e}my et al.\cite{Barthelemy+2022OD} first integrated SR and DA, in which forecasts are super-resolved by a neural network trained via supervised learning, and then assimilated by an ensemble Kalman filter. Although this approach can be easily incorporated into existing DA frameworks due to the separation of SR and DA, the non-local nature of SR may not be fully exploited in the assimilation process. Our previous study\cite{Yasuda+Onishi2023JAMES} proposed a different SRDA that performs end-to-end learning to jointly optimize SR and DA. This SRDA requires HR ground-truth data and does not provide a clear theoretical background for combining SR and DA. By contrast, the present SRDA does not rely on HR ground-truth data because it employs unsupervised learning based on the ELBO, and clarifies that the non-locality of SR benefits DA.

Other DA studies also use ELBOs.\cite{Ghimire+2017MICCAI, Xie+2023Authorea, Lafon+2023JAMES} The most distinct point from these studies is that they do not utilize VAEs. Indeed, the use of ELBOs is generally distinguished from the use of VAEs. When VAEs are used, ELBOs or underlying probability distributions often need to be designed specifically for VAEs.\cite{Kingma+2014, Sohn+2015NIPS} In our case, the approximate posterior $q\left(\bm{z}\mid \bm{x}, \bm{y}\right)$ is conditioned on both $\bm{x}$ and $\bm{y}$ in Eq. (\ref{eq:elbo-general-3}), which is essential for the encoder to perform assimilation (Sections \ref{subsec:3dvar-implementation-cvae} and \ref{subsec:srda-implementation-cvae}). Thus, one of our contributions is to provide a basis for the combination of ELBOs and VAEs for DA.

VAEs are already used for DA (see also Section \ref{sec:introduction}). Melinc and Zaplotnik\cite{Melinc+Zaplotnik2024QJQMS} performed dimensionality reduction using a VAE and then applied the 3D-Var in the latent space. Glyn-Davies et al.\cite{Glyn-Davies+2024JCP} employed a dynamical VAE (DVAE\cite{Girin+2021FTML}) to encode time series of observational data, which were then assimilated using a Kalman filter. Xiao et al.\cite{Xiao+2024arXiv} incorporated non-Gaussian background errors into the 3D-Var objective function by using the transformation with a VAE's decoder. The key distinction of our approach is that the CVAE itself performs assimilation (i.e., integrating observations into forecasts) using the non-locality of SR. This integration can be understood as the update by backward propagation of discrepancies from observations (Section \ref{subsec:srda-nonlinear-sr}).

Fundamentally, both DA and deep learning can be formulated using backpropagation with adjoint equations.\cite{Abarbanel+2018NC, Chen+2018NIPS, Geer2021PTRS} Using this analogy, recent studies have developed neural networks that update forecasts by propagating errors from observations along time.\cite{Lafon+2023JAMES, Fablet+2021, Beauchamp+2022GMDD} Our results reveal that SR models can act as error propagators for DA along the spatial dimensions, implying the possibility that neural networks act as propagators over the four dimensions (i.e., three spatial and one time dimensions).

\section{\label{sec:methods}Methods}

We evaluate the proposed SRDA method through numerical experiments similar to observing-system simulation experiments (OSSEs),\cite{Fletcher2023Chp25} which are commonly used in testing DA methods for the atmosphere and ocean. The ground truth (or the nature run) is generated by an HR fluid model, from which synthetic observations are computed. The SRDA employs an LR version of the fluid model, and the resultant forecast and analysis are compared with the ground truth.

A simple oceanic jet system was adopted for these experiments as a proof of concept for the SRDA. This jet exhibits the development of small-scale flow patterns and the sensitivity to initial conditions. SR can infer such small-scale patterns, while DA can suppress error growth due to the initial condition sensitivity. This jet system is suitable for assessing both aspects of SR and DA.

\subsection{\label{subsec:cfd-simulation} Fluid simulation}

An idealized barotropic ocean jet, proposed by David et al.,\cite{David+2017OM} was simulated in a two-dimensional periodic channel, using the same configuration as in Yasuda and Onishi.\cite{Yasuda+Onishi2023JAMES} The experimental setup is briefly explained here. See the above references for further details.

The governing equations in the periodic channel $(x,y)$ are as follows:
\begin{subequations}
    \label{eq:governing-eq-whole}
\begin{eqnarray}
  \frac{\partial \omega}{\partial t} + \bm{u \cdot \nabla} \omega + \beta \frac{\partial \psi}{\partial x} &=& - \kappa \omega - \nu \Delta^2 \omega - \frac{d\tau(y)}{dy}, \label{eq:vorticity-evolution} \\
  \Delta \psi &=& \omega, \label{eq:poisson-eq} \\
  \bm{u} &=& \left(-\frac{\partial \psi}{\partial y}, \frac{\partial \psi}{\partial x}\right), \label{eq:velocity-def}
\end{eqnarray}
\end{subequations}
where $x$ is the channel-wise direction and $y$ is the transverse direction ($x \in [0, 2\pi]$ and $y \in [0, \pi]$). Equation (\ref{eq:vorticity-evolution}) describes the evolution of vorticity $\omega$ over time $t$. The left-hand side represents the advection of $\omega$ and the beta effect,\cite{Vallis2017Chapter2} while the right-hand side comprises linear drag, hyperviscosity, and forcing due to zonal wind stress $\tau(y)$. Equations (\ref{eq:poisson-eq}) and (\ref{eq:velocity-def}) define the stream function $\psi$ and the velocity $\bm{u}$, respectively. All variables and parameters are non-dimensionalized following David et al.\cite{David+2017OM}

The parameters in Eq. (\ref{eq:vorticity-evolution}) were set as follows: $\beta = 0.1$, $\kappa = 1 \times 10^{-2}$, and $\nu = 1 \times 10^{-5}$. The wind stress was given by
\begin{equation}
    \tau(y) = \tau_0 \left[ {\text{sech}}^2\left(\frac{y - y_0}{\delta}\right) - c \right],
\end{equation}
where $\tau_0 = 0.3$, $y_0 = \pi/2$, $\delta = 0.4$, and $c$ was adjusted such that the integral of $\tau(y)$ was zero. This setup falls within the parameter regime of mixing barriers with strong eddies,\cite{David+2017OM} where coherent vortices persist in a statistically steady state (Section \ref{subsec:time-evolution-vorticity}).

The initial condition was a zonal jet superimposed with random perturbations. The zonal jet adopts the same shape as $\tau(y)$ and satisfies the Rayleigh-Kuo inflection-point criterion,\cite{Vallis2017Chapter9} which is a necessary condition for barotropic instability. Perturbations were added to the vorticity field for each wavenumber component, with phases and amplitudes drawn from Gaussian distributions. The data for deep learning were generated by varying these random perturbations.

Numerical integration was conducted using the modified Euler and pseudo-spectral methods at two spatial resolutions: LR and HR. Table \ref{table:cfd-simulation-config} provides detailed configurations for the LR and HR simulations. All simulations ran from $t = 0$ to $20$ with output time intervals of $0.25$. The CFD models using the LR and HR configurations are referred to as the LR and HR fluid models, respectively.

\begin{table}
    \caption{\label{table:cfd-simulation-config} Fluid simulation configurations at two different spatial resolutions: LR and HR. The cutoff wavenumber is the maximum wavenumber magnitude in the CFD model, which was determined by the specific condition to avoid aliasing errors in the pseudo-spectral method (i.e., the 2/3-rule).\cite{Canuto1988}}
    \centering
    \begin{ruledtabular}
    \begin{tabular}{l r r r}
        Name  & Grid size ($x \times y$) & Cutoff wavenumber & Integration time step \\
        \hline
        LR (low resolution) & $32 \times 16$  & 10 & $5 \times 10^{-4}$ \\
        HR (high resolution) & $128 \times 64$ & 42 & $\frac{1}{4} \times 5 \times 10^{-4}$ \\
    \end{tabular}
    \end{ruledtabular}
\end{table}

\subsection{\label{subsec:observations} Synthetic observations}

Synthetic observations were generated from the HR simulations. We consider an idealized point-wise observation network that imitates moored buoys in the ocean or balloon-borne radiosondes in the atmosphere. Observation points were subsampled at constant spatial intervals of $8 \times 8$ grid points, which means that 1.56\% ($=1/8^2$) of grid points were observed. To demonstrate that the SRDA can address changes in observation points, the subsampled points were varied in time by random shifts. For the unobserved points, Not a Numbers (NaNs) were assigned and later replaced with missing values in the preprocessing (Section \ref{subsec:neural-networks}). To emulate measurement errors, spatially independent Gaussian noise was added to the subsampled vorticity. This noise has a mean of 0 and a standard deviation of 0.1. This standard deviation is approximately 5\% of the spatio-temporal average of the absolute values of vorticity.

\subsection{\label{subsec:neural-networks} Neural networks}

We outline the network architectures of an SR model and a CVAE for the sequential SRDA. Hyperparameters are provided in the Appendix, and the full implementation is available at our Zenodo repository\cite{Yasuda+Onishi2024Code} (see Data Availability Statement). PyTorch 1.11.0\cite{PyTorch2019} was used for the implementation.

\begin{figure}
    \centering
    \includegraphics[width=16cm]{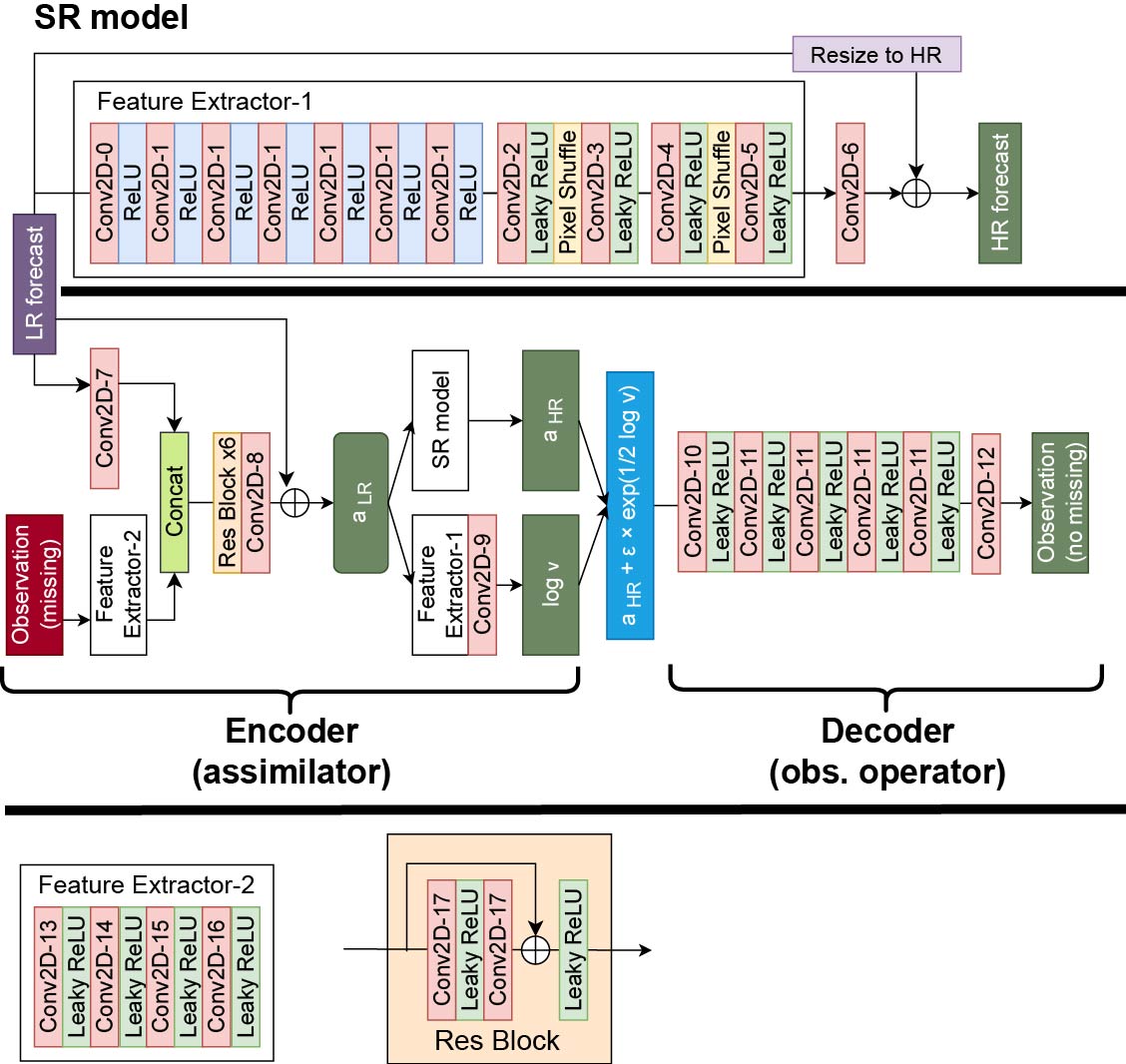}
    \caption{\label{fig:network_architecture} Network architectures of the SR model and the CVAE. The latter consists of an encoder and a decoder. The label ``Res Block'' refers to a residual block, and ``Res Block x6'' refers to the stacking of six Res Blocks. As in the typical implementation of VAEs,\cite{KingmaWelling2019FTML} the output of variance is regarded as $\ln V$ because neural networks can generally output both positive and negative values. Consequently, $\bm{a}_{\text{HR}} + \bm{\varepsilon}^{\text{T}} \exp(1/2 \log v)$ provides a realization of the true state $\bm{\hat{z}}$ in Eq. (\ref{eq:def-hat-z-srda}). All inputs and outputs consist only of vorticity (i.e., 1 channel).}
\end{figure}

The SR model architecture is shown in the top panel of Fig. \ref{fig:network_architecture}. This model resembles the neural network for SR known as VDSR, proposed by Kim et al.\cite{Kim+2016CVPR} The input is the two-dimensional vorticity field obtained from the LR fluid model, namely the LR forecast state, which is super-resolved to the HR. First, this input is nonlinearly transformed through a stack of convolution and ReLU layers. Upsampling is then executed by two blocks that include pixel shuffle layers.\cite{Shi+2016CVPR} Each block doubles the size of each dimension, resulting in a fourfold increase in resolution. Additionally, a skip connection\cite{He+2016CVPR} is employed, whereby the LR input, resized by bicubic interpolation, is added to the final output. This skip connection allows the SR model to capture the discrepancies between the LR input and the HR output.

The CVAE consists of an encoder and a decoder (Fig. \ref{fig:network_architecture}). The encoder performs SRDA and integrates the two inputs, an LR forecast and an HR observation, into the HR analysis. The HR observation is first downsampled through convolution to match its resolution with the LR (downsampled by a factor of 4). The effectiveness of separating forecast and observation inputs is shown in our previous study.\cite{Yasuda+Onishi2023JAMES} The LR forecast and the downsampled observation are then concatenated and passed through residual blocks to obtain the LR analysis $\bm{a}_{\text{LR}}$. This $\bm{a}_{\text{LR}}$ is super-resolved by the trained SR model to the HR analysis $\bm{a}_{\text{HR}}$. The variance for $\bm{a}_{\text{HR}}$ [$V$ in Eq. (\ref{eq:def-variance-srda})] is also estimated by the encoder. Although $\bm{a}_{\text{HR}}$ and $V$ can be treated independently in the encoder (Section \ref{subsec:srda-implementation-cvae}), we compute $V$ from $\bm{a}_{\text{LR}}$ to reduce the number of parameters, using the feature extractor in the trained SR model (i.e., Feature Extractor-1 in Fig. \ref{fig:network_architecture}).

The decoder acts as an observation operator and reconstructs the observations from $\bm{\hat{z}}$. This operation comprises convolution and Leaky ReLU.\cite{Maas+2013ProcICML} We confirmed that the decoder can be replaced with a subsampling operation (not a neural network), though this replacement slightly decreases the inference accuracy. This result is due to the experimental setup where the synthetic observations were generated by subsampling (Section \ref{subsec:observations}).

In preprocessing, the vorticity $\omega$ at each grid point is transformed by
\begin{equation}
    {\text{clip}}_{[0, 1]}\left(\frac{\omega - m_\omega}{s_\omega}\right). \label{eq:preprocess}
\end{equation}
The clipping function, ${\text{clip}}_{[0,1]}(z) = \min\{1, \max\{0, z\}\}$ ($z \in \mathbb{R}$), restricts the value range to the interval $[0, 1]$. The parameters $m_\omega$ and $s_\omega$ are determined such that 99.9\% of the vorticity values fall within the $[0,1]$ range. After applying Eq. (\ref{eq:preprocess}), zero is regarded as a missing value and is assigned to the HR grid points without observations. This clear interpretation of zero facilitates feature extraction from the input data.\cite{Yasuda+2023BAE}

\subsection{\label{subsec:train-test-methods} Training and testing methods for the neural networks}

\subsubsection{Training method}

\begin{figure}
    \centering
    \includegraphics[width=13cm]{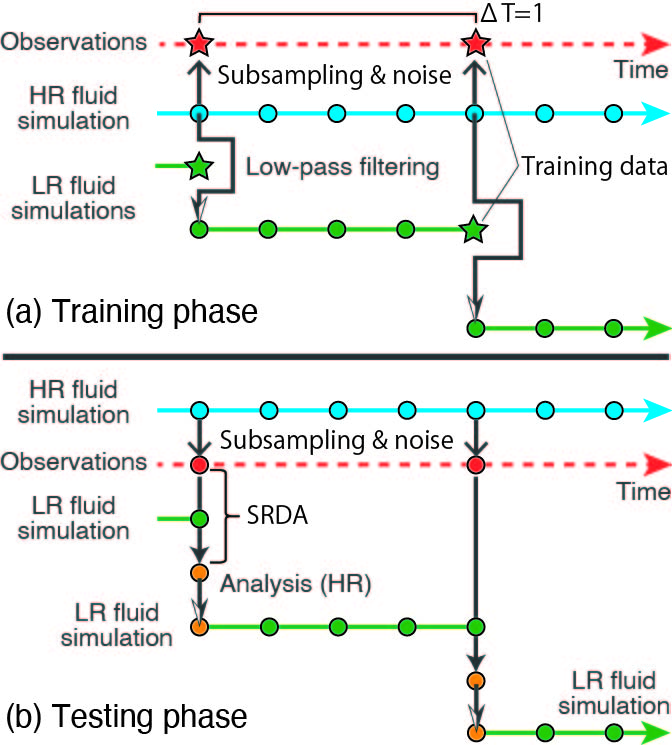}
    \caption{\label{fig:train_test_methods} Schematics of time series for (a) training and (b) testing phases. The assimilation interval is set to $\Delta T=1$, while the output interval (i.e., the forecast interval) is $\Delta T/4 = 0.25$.}
\end{figure}

We adopted an offline strategy in which training data were pre-generated [Fig. \ref{fig:train_test_methods}(a)]. The training data consist of pairs of HR observations and LR forecasts, as discussed in Sections \ref{subsec:loss-for-sr-models} and \ref{subsec:srda-implementation-cvae}. The initial conditions for the LR fluid model were obtained by applying a low-pass filter to the HR simulation results in the wavenumber domain. This filtering completely removes all high-wavenumber components using discrete Fourier transform. After time integration over $\Delta T$ ($= 1$), the resultant LR forecasts were used in training. This $\Delta T$ equals the assimilation interval, which is a reasonable timescale because $\Delta T = 1$ is approximately half of the advection time of the zonal jet. The corresponding observations were obtained by subsampling. Figure \ref{fig:train_test_methods}(a) denotes pairs of the LR forecasts and HR observations as stars. By varying the random initial perturbations (Section \ref{subsec:cfd-simulation}), we conducted 4,500 simulations; 70\% were used as training data and the remaining 30\% as validation data for hyperparameter tuning.

In principle, HR simulations are not required because the training data include LR forecasts rather than HR results. These LR forecasts are obtained by numerical integration in which the analysis states are used as initial conditions. In offline training, however, such an analysis is unavailable because the CVAE is not trained yet. Thus, the filtered HR results were used as the initial conditions. In realistic applications, these initial conditions can be obtained from existing assimilated data,\cite{Hersbach+2020QJRMS, Kosaka+2024JMSJ} or online training would be effective,\cite{Sahoo+2018IJCAI,Rasp2020GMD} where training and data generation are performed alternately or in parallel.

The SR model was trained, followed by training the CVAE. For both trainings, the Adam optimizer\cite{Kingma+2015ICLR} was used with a learning rate of $1 \times 10^{-4}$ and a mini-batch size of 128. Each training was terminated using early stopping with a patience parameter of 50 epochs. During the CVAE training, we fixed all parameters in the SR model. Note that VAEs typically show stable training and inference characteristics\cite{Kingma+2014ICLR} and we confirmed this behavior in our experiments (not shown). The loss functions of the SR model and the CVAEs are $l_{\text{SR}}$ in Eq. (\ref{eq:loss-sr}) and $l_{\text{SRDA}}$ in Eq. (\ref{eq:elbo-srda}), respectively. The parameters in $l_{\text{SRDA}}$ were set to $r = 1.2 \times 10^{-5}$ and $b = 1.0 \times 10^{-3}$. The value of $r$ was determined by the variance of the measurement errors, while $b$ was determined through grid search to balance the reconstruction error and the KL divergence in the ELBO. Without this balancing, the performance of the CVAE would deteriorate. This phenomenon frequently occurs in VAEs,\cite{Higgins+2017ICLR, Alemi+2018PMLR} and the loss balancing can be optimized during training.\cite{AspertiTrentin2020IEEE} For simplicity, however, this advanced technique was not employed in our study.

\subsubsection{Testing method}

The sequential SRDA was tested using the method described in Section \ref{subsec:sequentiaSRDA-L}. Figure \ref{fig:train_test_methods}(b) shows a schematic of the test phase. The only difference between Figs. \ref{fig:sequential_SRDA} and \ref{fig:train_test_methods}(b) is that observations are generated by subsampling the HR simulation results. Due to this difference, the HR results are regarded as the ground truth, and the SRDA is evaluated by comparing its inferences to the ground truth. We newly conducted 500 HR fluid simulations. For each case, the LR fluid model was initialized with the low-pass filtered HR vorticity at $t=0$; the SRDA was subsequently conducted until $t = 20$ with the assimilation interval $\Delta T = 1$. Without SRDA, the vorticity evolution would significantly differ from the ground truth, not only due to differences in the initial condition (i.e., all high-wavenumber components are completely removed) but also due to the difference in the fluid model resolution (i.e., LR versus HR).

Two test metrics were used to assess pixel-wise accuracy and pattern consistency, both of which measure deviations from the ground truth. These two types of metrics are frequently used in SR studies\cite{Lepcha+2023IF} because they complement each other. The pixel-wise accuracy quantifies localized differences but may overestimate discrepancies when spatial patterns are correctly inferred with a slight shift. The pattern consistency mitigates this effect by evaluating structural similarity between inferred and ground-truth patterns, although it may overlook significant pixel-wise deviations.

The mean absolute error (MAE) ratio quantifies pixel-wise errors.
\begin{equation}
    \text{MAE ratio} = \frac{\sum_{i} |\omega_i - \hat{\omega}_i|}{\sum_{i} |\omega_i|}, \label{eq:mae-ratio}
\end{equation}
where $\omega_i$ denotes the ground-truth vorticity at the $i$-th grid point, $\hat{\omega}_i$ is the corresponding inference, and the summation is taken over all grid points.

The mean structural similarity index measure (MSSIM) loss evaluates the consistency of spatial patterns:\cite{Wang+2004IEEE}
\begin{eqnarray}
    \text{MSSIM loss} &=& 1 - {\text{MSSIM}}, \nonumber \\ &=& 1 - \sum_i \frac{\left(2\mu_{i}\hat{\mu}_{i} + {\text{C}_1}^2 \right) \left(2\gamma_{i} + {\text{C}_2}^2 \right)}{\left(\mu_{i}^2 + \hat{\mu}_{i}^2 + {\text{C}_1}^2\right) \left( \sigma_{i}^2 + \hat{\sigma}_{i}^2 + {\text{C}_2}^2 \right)}, \label{eq:mssim}
\end{eqnarray}
where ${\text{C}_1} = 0.01$, ${\text{C}_2} = 0.03$, $\mu_{i}$ and $\sigma_{i}^2$ are the mean and variance of the ground truth, respectively, $\hat{\mu}_{i}$ and $\hat{\sigma}_{i}^2$ are the corresponding quantities for the inference, and $\gamma_{i}$ is the covariance between the ground truth and the inference. These variables in the summation are calculated locally in space using a Gaussian filter. The MSSIM loss takes a value greater than or equal to zero; a smaller value indicates spatial patterns more similar to the ground truth. A detailed discussion of MSSIM can be found in Wang et al.\cite{Wang+2004IEEE}

The MAE ratio and MSSIM loss were calculated at each time step in each test simulation. The values averaged across all test simulations are referred to by the same names. Both MAE ratio and MSSIM loss are collectively called the test errors. We discuss the results from the test data using these metrics in Section \ref{sec:results-discussion}.

\subsection{\label{subsec:enkf} Baseline model using Ensemble Kalman Filter (EnKF)}

As a baseline for comparison with the SRDA, we employed an ensemble Kalman filter (EnKF)\cite{Evensen1994JGRO} for the LR fluid model, using the perturbed observation method.\cite{Burgers+1998MWR} The choice of EnKF was motivated by its ability to estimate uncertainty through ensemble simulations. The uncertainties estimated by the EnKF and SRDA are compared in Section \ref{subsec:accuracy-srda-analysis}.

Assimilation was conducted in the HR space after the LR forecast was super-resolved to the HR using bicubic interpolation, with the assimilation interval $\Delta T$ ($=1$) being the same as in the SRDA. Barth\'{e}l\'{e}my et al.\cite{Barthelemy+2022OD} argued that performing the EnKF in the HR space can yield more accurate inferences than those in the LR space when observations are defined at HR grid points. Furthermore, this approach clarifies the comparison with the SRDA because both EnKF and SRDA employ the LR fluid model and perform DA in the HR space.

The background error covariance matrices were spatially localized using a function proposed by Gaspari and Cohn.\cite{Gaspari+1999QJRMS} These error covariances were inflated by adding Gaussian noise to the analysis before the forecast process.\cite{Whitaker+2008MWR} This Gaussian noise incorporates spatial correlation by estimating the covariance from the training data at each time step. The initial perturbations for creating ensemble members were also drawn from this Gaussian distribution.

Hyperparameters for the EnKF were adjusted using the training data to minimize the MAE. Specifically, the following parameters were tuned: the number of ensemble members (set to 300 after tuning), the amplitude of the initial perturbation for generating ensemble members, the amplitude of the additive inflation, the localization radius, and the amplitude for perturbing observations. The full implementation of the EnKF is also available at our Zenodo repository\cite{Yasuda+Onishi2024Code} (see Data Availability Statement).

\section{\label{sec:results-discussion}Results and discussion} 

\subsection{\label{subsec:time-evolution-vorticity}Time evolution of vorticity in the jet system}

We first describe the time evolution of HR vorticity in the jet system and confirm the sensitivity to initial conditions (i.e., a positive Lyapunov exponent).

\begin{figure}
    \centering
    \includegraphics[width=16cm]{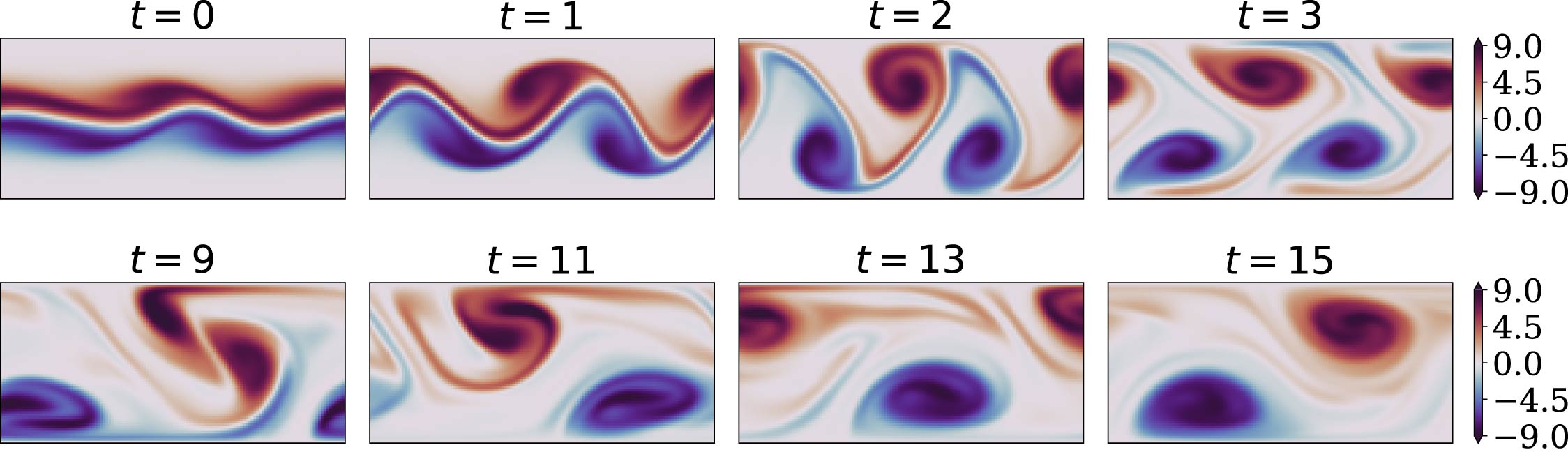}
    \caption{\label{fig:vorticity-evolution} Typical vorticity evolution from the HR ground-truth data.}
\end{figure}

Figure \ref{fig:vorticity-evolution} shows a typical evolution of HR vorticity. Initially ($t=0$), an unstable jet exists around the $y$ center with small perturbations. The jet starts meandering and breaks down into multiple vortices, accompanied by the development of fine filaments ($1 \lesssim t \lesssim 4$). Subsequently, vortex merging occurs, intensifying small-scale structures again ($4 \lesssim t \lesssim 12$). Finally, a wavenumber-1 structure continues to propagate westward ($12 \lesssim t$). This final state is statistically steady and can be understood by Rossby wave dynamics.\cite{David+2017OM}

The maximum Lyapunov exponent $\lambda$ is positive: $\lambda = 3.11$. This indicates that the temporal evolution of vorticity exhibits initial-value sensitivity, likely due to the barotropic instability. Without DA, the deviation from the ground truth continues to increase over time, as shown in Section \ref{subsec:accuracy-srda-forecast}.

\subsection{\label{subsec:accuracy-sr-model}Accuracy of SR inference}

\begin{figure}
    \centering
    \includegraphics[width=16cm]{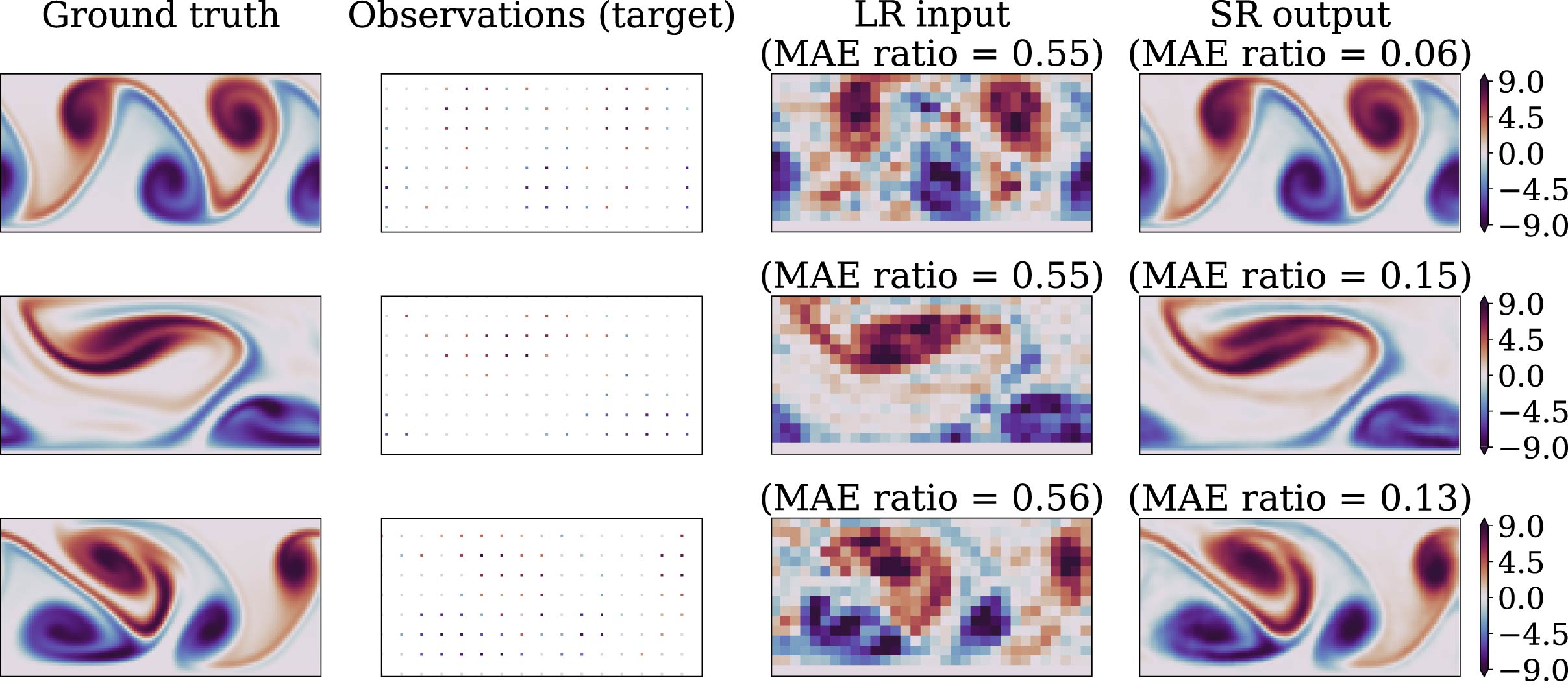}
    \caption{\label{fig:sr_snapshots} Vorticity snapshots for the ground truth, observations, LR input, and SR output. The SR model was trained using the observation data as target. The input data are not indexed by time because each sample was randomly selected for testing without simulating time evolution.}
\end{figure}

We show the accuracy of inference by the SR model, as the SRDA is based on this model. Figure \ref{fig:sr_snapshots} shows three sets of snapshots for the ground truth, observations, LR input, and SR output. The SR model reproduces the fine filaments and the small-scale structures within vortices. These patterns are not evident in the LR inputs or observations.

Flow-field reconstruction has been extensively studied using neural networks.\cite{Francesc+2020JASA, Maulik+2020PRF, Gundersen+2021PF, Wang+2022MWR} In these studies, inputs are incomplete, such as spatially sparse sensor measurements. Neural networks are then trained by comparing their outputs to complete target data. Our problem setup is opposite to that of these studies: in our case, the input is complete, while the target (i.e., the observation) is incomplete.

The success of SR in our study is attributable to two factors: (i) the point-wise nature of the loss function and (ii) the weight sharing of convolution. The loss function $l_{\text{SR}}$ in Eq. (\ref{eq:loss-sr}) operates at individual grid points, and only the non-missing observed points contribute to $l_{\text{SR}}$. This point-wise nature enables backpropagation even when the target data are missing. Furthermore, since convolution shares its kernel across space (known as weight sharing),\cite{Aloysius+Geetha2017IEEE} the efficient learning was realized using the small amount of non-missing observational data.

Although the inference appears to be accurate, the SR model cannot adequately correct forecast errors. One might think that the SR output could be used as the analysis due to its high accuracy (Fig. \ref{fig:sr_snapshots}). However, test errors accumulated over time when the downsampled SR output was used as the initial condition for the LR fluid model. This result indicates that observations need to be assimilated into forecasts.

The trained SR model makes non-local inference. This non-locality is essential for assimilating observations, as discussed in Section \ref{subsec:srda-nonlinear-sr}. We use two indicators for interpreting the SR model: gradients and local attribution maps (LAMs).\cite{Gu+Dong2021CVPR} Gradients are expressed as $\partial \mathsf{F}(\bm{x}) / \partial \bm{x}$, where $\mathsf{F}$ is the trained SR model. Such a simple gradient is sometimes not a good indicator for interpreting how a model output depends on its input due to nonlinearity.\cite{Gu+Dong2021CVPR, Sundararajan+2017ICML} LAM, proposed by Gu and Dong,\cite{Gu+Dong2021CVPR} combines gradient methods with path integral techniques for interpreting behaviors of nonlinear SR models.

\begin{figure}
    \centering
    \includegraphics[width=16cm]{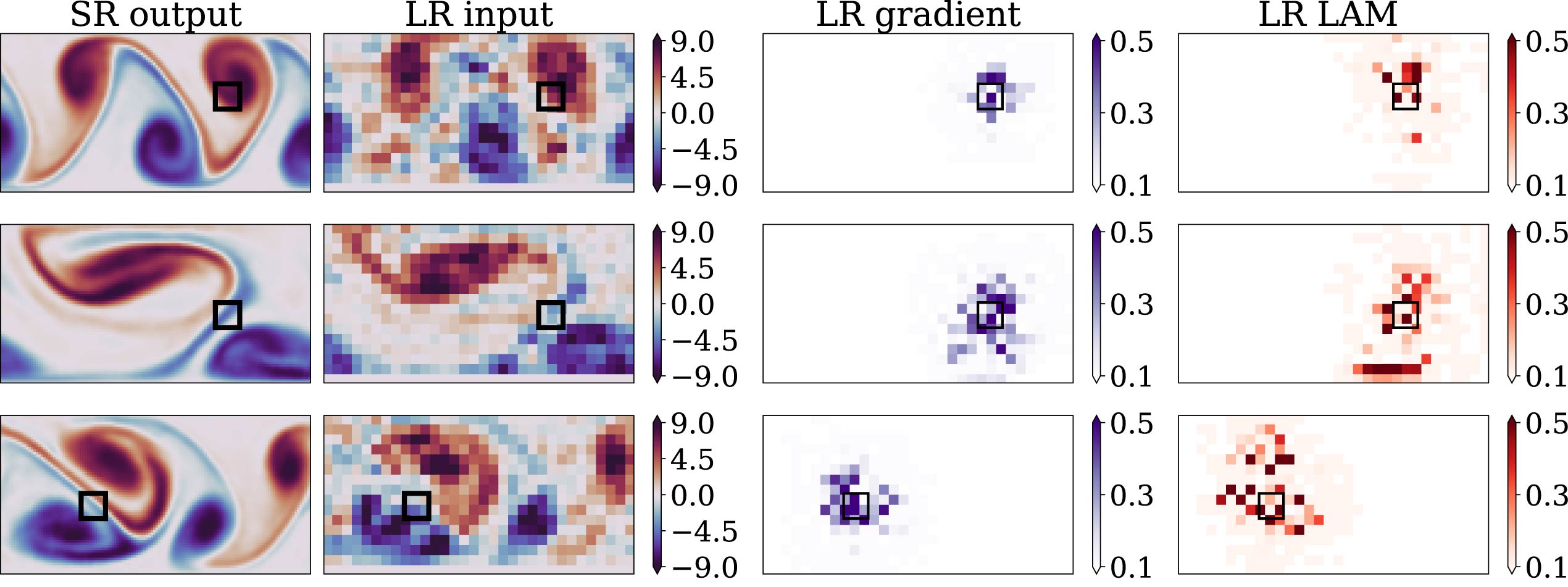}
    \caption{\label{fig:sr_lam} Attribution maps for SR. The left two columns, SR output and LR input, are the same as in Fig. \ref{fig:sr_snapshots}, but shown here for comparison. Local attribution maps (LAMs)\cite{Gu+Dong2021CVPR} were calculated for the small square regions of $11\times 11$ (i.e., black boxes in the figure). These boxes were specified on the HR grids, so they are not necessarily aligned with the LR grids. The gradients and LAMs were normalized by their maximum values, with ranges between 0 and 1, following Gu and Dong.\cite{Gu+Dong2021CVPR}}
\end{figure}

Figure \ref{fig:sr_lam} shows gradient and LAM fields for the three SR samples previously discussed in Fig. \ref{fig:sr_snapshots}. We focus on the square regions (black boxes) in the figures. The gradients and LAMs indicate that colored points in the LR inputs contribute to the HR outputs in these boxes. If there are observed values in these boxes, the colored points in the inputs are updated by minimizing the loss function (Section \ref{subsec:srda-nonlinear-sr}). Thus, observations have a non-local influence through the trained SR model. This is an essential reason for the success of the proposed SRDA. Interestingly, the LAMs tend to take larger values over wider regions, compared to the gradients (Fig. \ref{fig:sr_lam}), suggesting the nonlinearity of the SR model. For instance, in the first row of the figure, the LAM has a larger value at one grid point below the black box, matching the position of the vortex filament.

\subsection{\label{subsec:accuracy-srda-forecast}Accuracy of LR forecast}

We demonstrate that the sequential SRDA can make more accurate inferences than the EnKF. The LR forecast is discussed here, and the HR analysis is examined next.

\begin{figure}
    \centering
    \includegraphics[width=16cm]{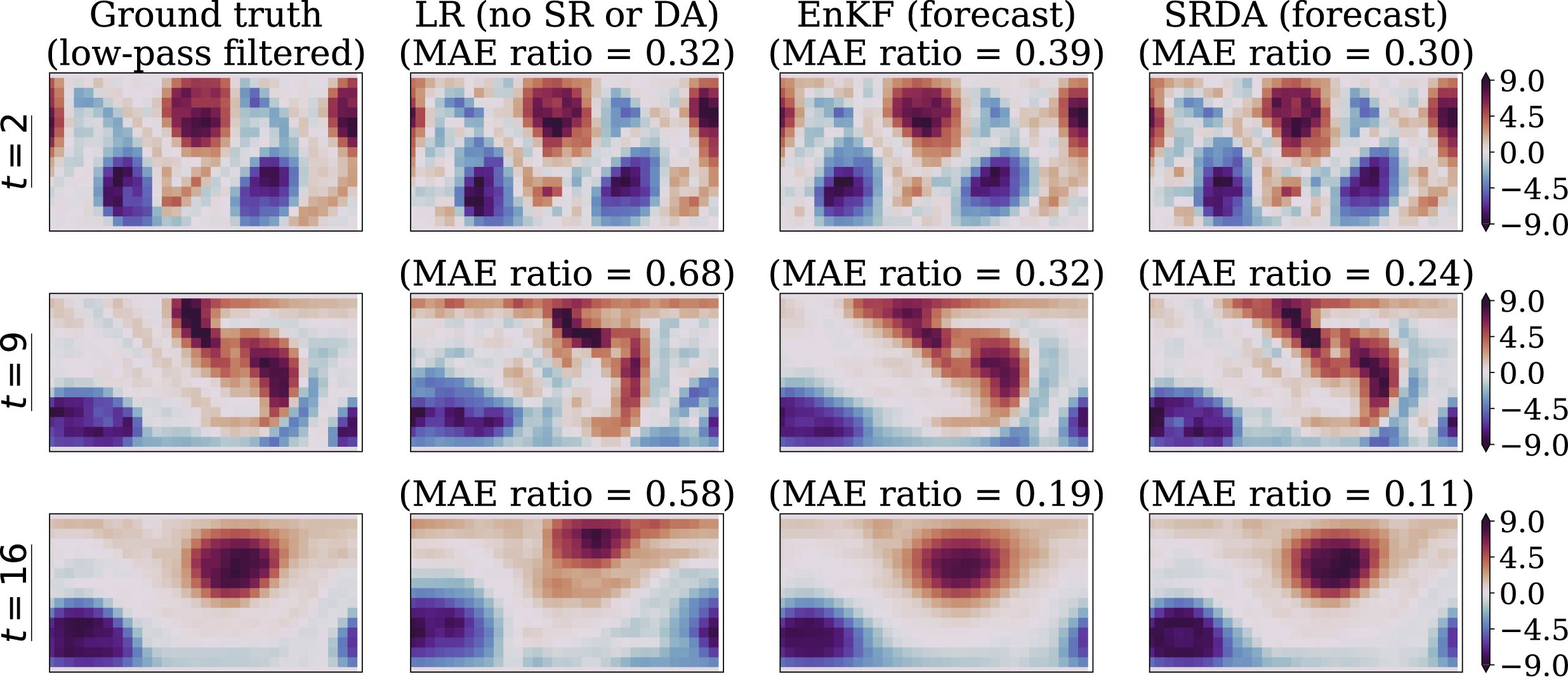}
    \caption{\label{fig:srda_forecast_snapshots} Vorticity snapshots of the ground truth and forecasts for the EnKF and SRDA at $t=2$, 9, and 16. For comparison, the figure also shows LR simulation results without SR or DA. A low-pass filter was applied to the ground-truth snapshots to match their resolution to the LR.}
\end{figure}

Figure \ref{fig:srda_forecast_snapshots} shows snapshots of the ground truth and forecasts for the EnKF and SRDA. To match the resolution, a low-pass filter was applied to the HR ground-truth vorticity. The figure also includes LR simulation results without SR or DA. During the jet collapse ($t = 2$), differences in the inferences are not obvious, likely because the LR fluid model cannot well emulate fine filaments. During the vortex merging ($t = 9$), the SRDA most accurately describes the vortex-split distribution near the center. The EnKF also captures this pattern, but the SRDA inference is closer to the ground truth. Without SR or DA, the vortex-split structure is not reproduced. In the steady state ($t = 16$), both EnKF and SRDA accurately infer the vorticity phase. Without SR or DA, the vortex shapes and locations still differ from the ground truth.

\begin{figure}
    \centering
    \includegraphics[width=14.25cm]{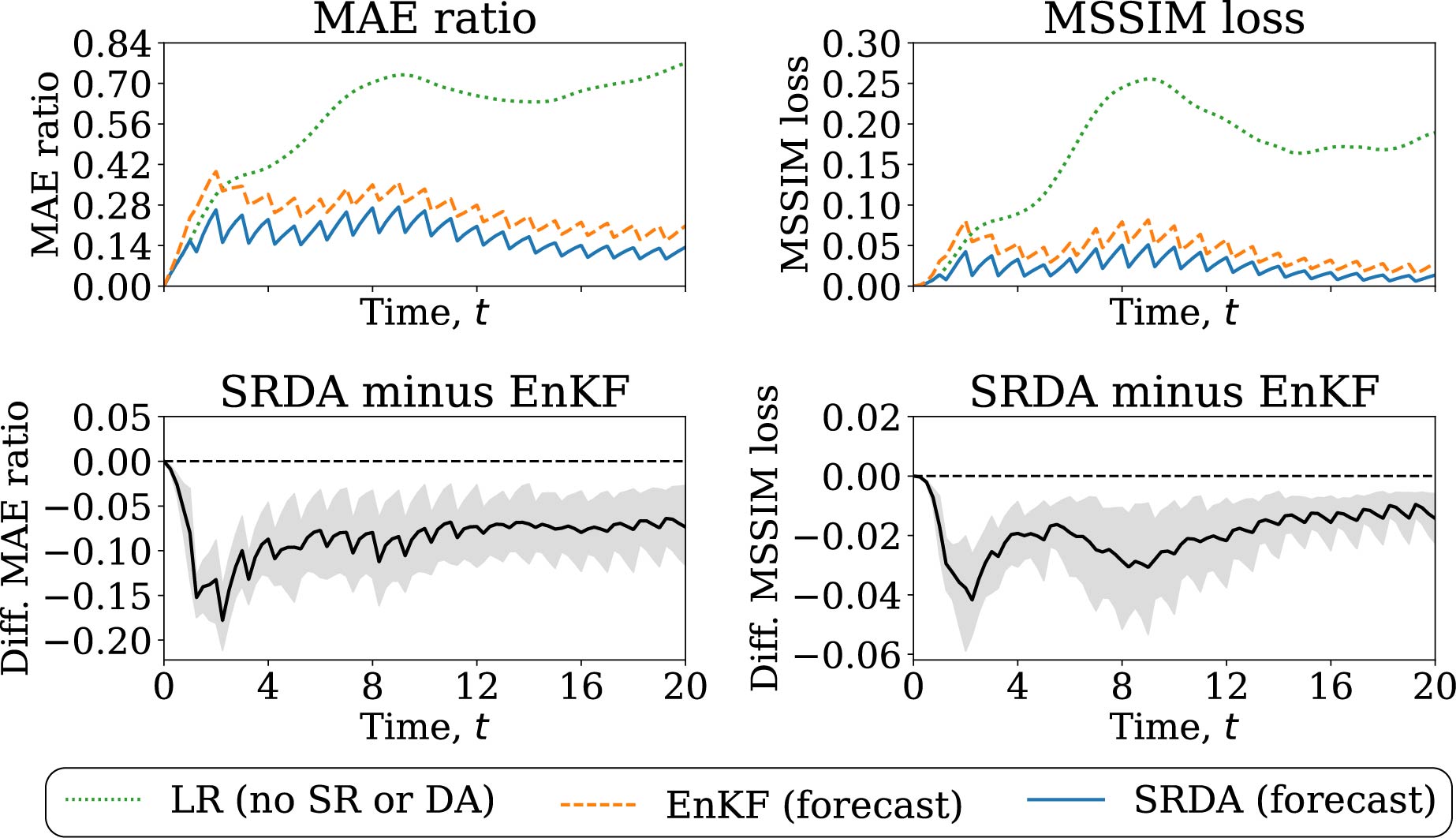}
    \caption{\label{fig:srda_forecast_time_series} Time series of the MAE ratio and MSSIM loss for the forecast states in the 500 test simulations. The top panels show the mean test errors. The bottom panels show the differences between the SRDA and EnKF, where the gray areas cover the 10th to 90th percentile values and the solid lines indicate the mean differences.}
\end{figure}

Figure \ref{fig:srda_forecast_time_series} shows the time series of the mean values for the MAE ratio and MSSIM loss. Without SR or DA (green dotted lines), the test errors increase over time due to the sensitivity to initial conditions (Section \ref{subsec:time-evolution-vorticity}). In contrast, this error growth is suppressed by the EnKF (orange dashed lines) and the SRDA (blue solid lines). The error reduction occurs every time unit since the assimilation interval was set to 1. When smoothed over time, the time series of the EnKF and SRDA have two peaks, corresponding to the two developments of small-scale patterns during the jet collapse ($t \sim 2$) and the vortex merging ($t \sim 9$). The bottom panels in Fig. \ref{fig:srda_forecast_time_series} show the differences in the test errors between the EnKF and SRDA, where the gray areas represent the 10th to 90th percentile values. The test errors of the SRDA are significantly smaller than those of the EnKF. Thus, we conclude that the SRDA forecast outperforms the EnKF in terms of pixel-wise accuracy and pattern consistency.

\begin{figure}
    \centering
    \includegraphics[width=16cm]{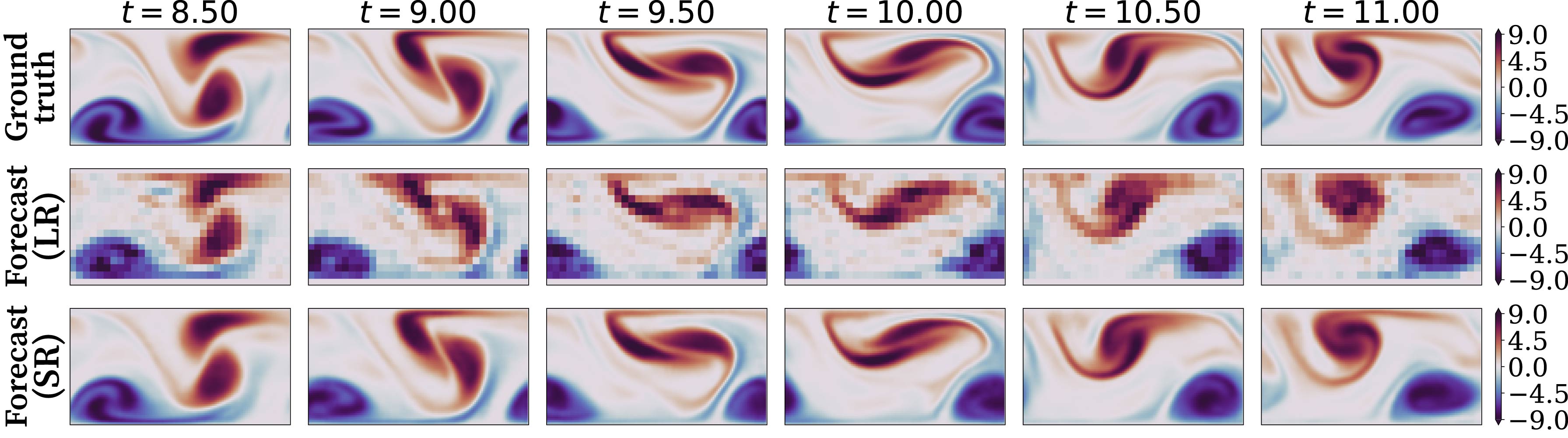}
    \caption{\label{fig:sr_snapshots_in_srda} Typical vortex merging in the ground truth, the LR forecasts, and the super-resolved forecasts.}
\end{figure}

In the proposed SRDA, the LR forecast can be super-resolved to the HR at an arbitrary time using the SR model (Section \ref{subsec:sequentiaSRDA-L}). Figure \ref{fig:sr_snapshots_in_srda} shows a typical vortex merging between $t=8.5$ and 11.0 from the ground truth, the LR forecast, and the super-resolved forecast. The merging of two vortices is not obvious in the LR, but is well captured in the super-resolved forecast.

\subsection{\label{subsec:accuracy-srda-analysis}Accuracy of HR analysis}

\begin{figure}
    \centering
    \includegraphics[width=13.75cm]{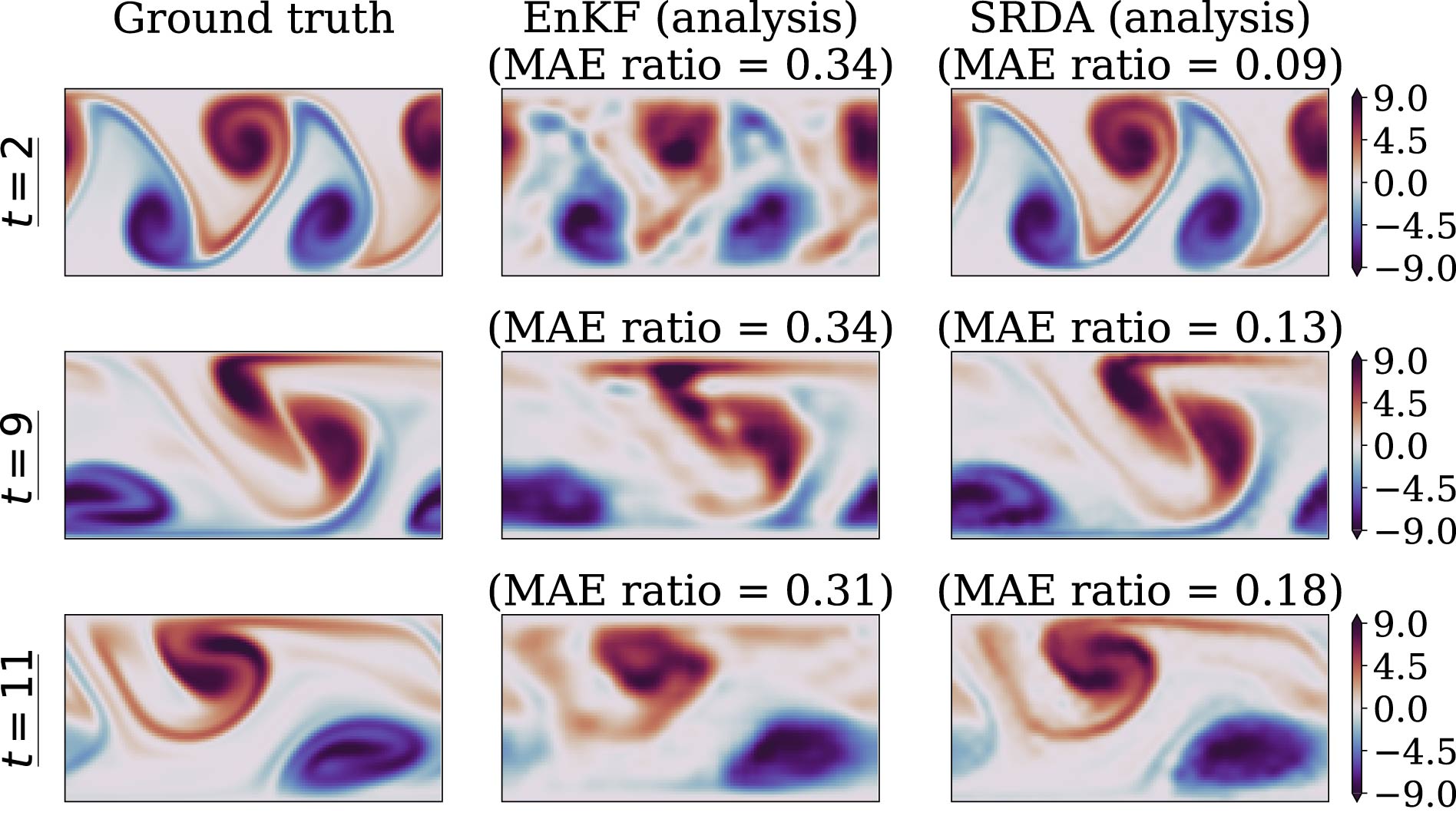}
    \caption{\label{fig:srda_analysis_snapshots} Vorticity snapshots of the ground truth and analyses for the EnKF and SRDA at $t=2$, 9, and 11.}
\end{figure}

Figure \ref{fig:srda_analysis_snapshots} shows snapshots of the ground truth and analyses. During the jet collapse ($t = 2$), the filament structure appears much more distinct in the SRDA analysis than in the EnKF. Furthermore, during the vortex merging ($t = 9$ and 11), the SRDA infers sharper vortex shapes and clearer internal structures than the EnKF.

\begin{figure}
    \centering
    \includegraphics[width=14.25cm]{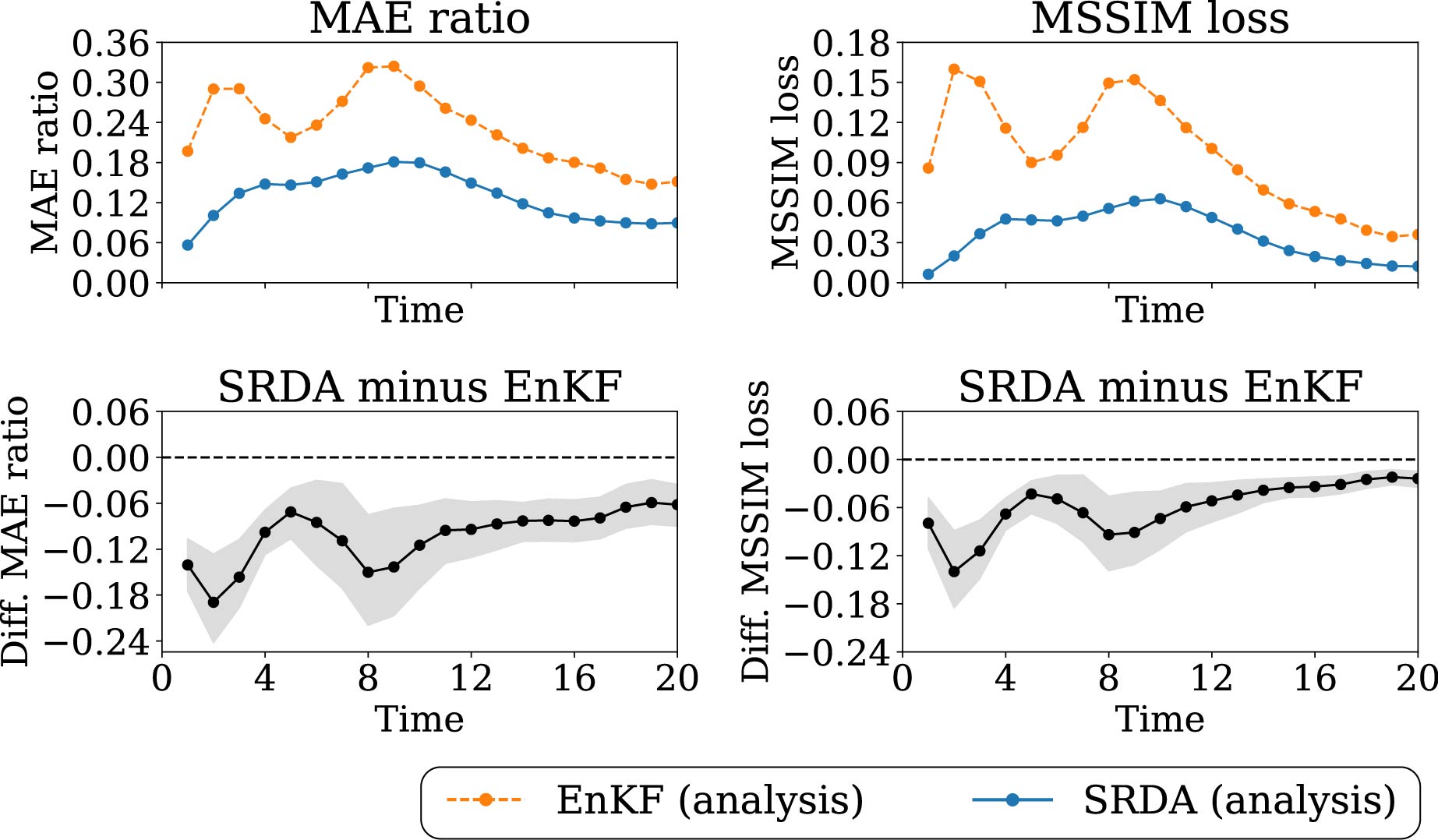}
    \caption{\label{fig:srda_analysis_time_series} Time series of the MAE ratio and MSSIM loss for the analysis states in the 500 test simulations. These states were inferred only at the analysis time steps, denoted by dots. The top panels show the mean test errors. The bottom panels show the differences between the SRDA and EnKF, where the gray areas cover the 10th to 90th percentile values and the solid lines indicate the mean differences.}
\end{figure}

Figure \ref{fig:srda_analysis_time_series} shows the time series of the MAE ratio and MSSIM loss, as in Fig. \ref{fig:srda_forecast_time_series}. The dots represent the assimilation time steps, with assimilation intervals at 1. The EnKF and SRDA analyses are inferred only at these times. The EnKF time series (orange dashed lines) show two peaks, corresponding to the two developments of small-scale patterns during the jet collapse ($t \sim 2$) and the vortex merging ($t \sim 9$). These peaks are less apparent in the SRDA time series (blue solid lines). The bottom panels in Fig. \ref{fig:srda_analysis_time_series} indicate that the test errors of the SRDA are significantly smaller than those of the EnKF. Thus, we conclude that the SRDA analysis is more accurate than the EnKF analysis.

We compare uncertainties estimated by the EnKF and SRDA, using standard deviation (SD) as an indicator. For the EnKF, SD at each grid point was calculated from all ensemble members (i.e., 300 members). For the SRDA, SD was directly calculated by taking the square root of the estimated variances in Eq. (\ref{eq:def-variance-srda}).

\begin{figure}
    \centering
    \includegraphics[width=16cm]{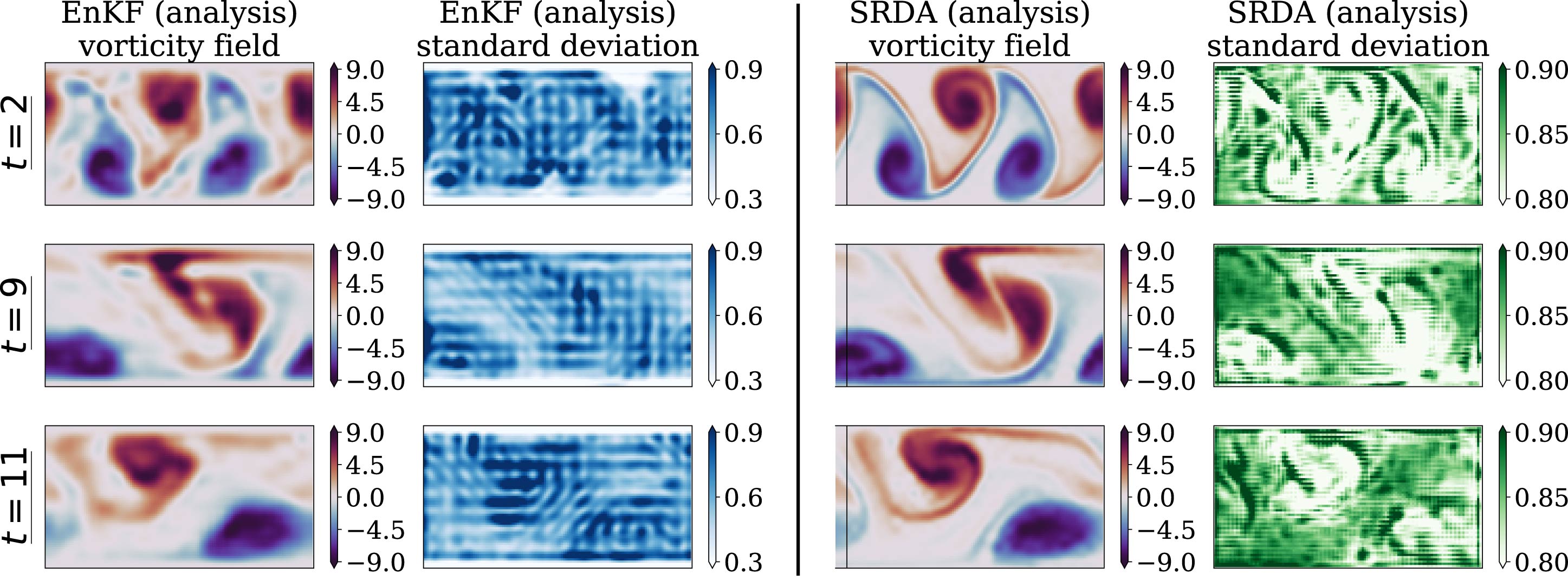}
    \caption{\label{fig:uncertainties_enkf_srda} Vorticity snapshots of the analyses and their standard deviations (SDs) for the EnKF and SRDA at $t=2$, 9 and 11. The analyses shown here are the same as in Fig. \ref{fig:srda_analysis_snapshots}.}
\end{figure}

Figure \ref{fig:uncertainties_enkf_srda} compares the analyses and their SDs for the EnKF and SRDA. The SRDA SD tends to be large around vortex filaments or vortex edges. This tendency appears to be most evident during the jet collapse ($t=2$). This result is reasonable because vortex locations are generally uncertain, and small changes in their locations lead to large differences in vorticity. The EnKF SD also exhibits a similar tendency; however, the SD distribution is blurred, likely reflecting the blurred vortex patterns. The SD value ranges also differ: the EnKF SD is between 0.3 and 0.9, while the SRDA SD is between 0.8 and 0.9. Compared to the EnKF, the SRDA tends to estimate larger SD in regions where vorticity is close to zero, especially at $t=9$ and 11 in Fig. \ref{fig:uncertainties_enkf_srda}. Although the true value of SD is unknown, our results suggest that the SRDA tends to estimate larger SD in low-vorticity regions. This tendency may be due to a simplified variance estimation in the present CVAE (Section \ref{subsec:neural-networks}), where the variances are computed using the SR model from the LR analysis. The SRDA demonstrated a potential for uncertainty quantification without relying on ensemble calculation; however, further model development is necessary. Recent studies demonstrate that VAEs can learn error distributions of model forecasts.\cite{Groom2021QJRMS, Yang+2021JCP, Grooms+2023QJRMS, Xiao+2024arXiv} Their model architectures may be effective for the SRDA.

\subsection{\label{subsec:computation-time-srda}Computation time of SRDA}

The computational time of the SRDA is much less than that of the EnKF. The elapsed times for the simulations from $t = 0$ to 20 were measured using a single process on a CPU (Intel Xeon Gold 6326). For the SRDA, in addition to the assimilation by the encoder, the LR forecast was super-resolved by the SR model at every output time step ($\Delta t = 0.25$). The average wall time was 21.3 s for the SRDA, considerably less than 205.7 s for the EnKF. This difference is mainly due to the calculation of the ensemble evolution, which is not required in the SRDA.

\section{\label{sec:conclusions}Conclusions}

This study proposed a theory of unsupervised SRDA using the CVAE. By deriving an ELBO for DA, we showed that the theory is an extension of the 3D-Var (Section \ref{sec:da-elbo}). Since ELBOs generally serve as the loss function for CVAEs, the assimilation is feasible using the CVAE. In this theory, it is essential to use the non-local SR operation, which makes background covariances learnable from data (Section \ref{sec:srda-elbo}). The proposed SRDA assimilates HR observations into LR forecasts without running HR fluid models. The effectiveness of the proposed method was evaluated via numerical experiments using an idealized barotropic ocean jet system (Sections \ref{sec:methods} and \ref{sec:results-discussion}). Compared to the EnKF, the SRDA demonstrated its ability to obtain accurate HR inferences within shorter computational times.

There are several future research directions for unsupervised SRDA. The major limitation of the proposed theory is the Gaussian assumption Eq. (\ref{eq:gaussian-whole-srda}), which suggests that the accuracy in inference would be reduced if a system exhibited strong non-Gaussianity. One method to extend the theory to non-Gaussian systems is combining CVAEs with normalizing flows\cite{Rezende+2015PMLR, Kobyzev+2021IEEE} to learn non-linear transformations that lead to non-Gaussian distributions. The proof of concept here has only examined the linear observation operator, although the proposed SRDA theoretically allows learning of nonlinear observation operators. Furthermore, advanced training methods, such as online training\cite{Sahoo+2018IJCAI,Rasp2020GMD} and loss balancing for VAEs,\cite{AspertiTrentin2020IEEE} will be important to develop SRDA. As for generalizability, we have not evaluated to what extent the CVAE remains effective when the statistical properties of the input data differ. These aspects should be addressed in a more realistic experimental setup. Indeed, the fluid system used here is simple compared to actual atmospheric and oceanic systems. It is necessary to evaluate the proposed SRDA for more complex systems.

\begin{acknowledgments}
This work used computational resources of the TSUBAME3.0 supercomputer provided by the Institute of Science Tokyo through the HPCI System Research Project (Project ID: hp220102) to conduct the CFD simulations. The deep learning was performed on the Earth Simulator system (Project IDs: 1-23007 and 1-24009) at the Japan Agency for Marine-Earth Science and Technology (JAMSTEC). This paper is based on results obtained from a project, JPNP22002, commissioned by the New Energy and Industrial Technology Development Organization (NEDO).
\end{acknowledgments}

\section*{Author Declarations}
\subsection*{Conflict of Interest}
The authors have no conflicts to disclose


\section*{Data Availability Statement}

The data and source code that support the findings of this study are openly available at our Zenodo repository (\url{https://doi.org/10.5281/zenodo.13958042}), reference number \onlinecite{Yasuda+Onishi2024Code}. These files are also available at our GitHub repository (\url{https://github.com/YukiYasuda2718/srda-cvae/releases/tag/v0.2.1}).

\appendix*

\section{Hyperparameters for the SR model and the CVAE}

We list the hyperparameters of the SR model and the CVAE shown in Fig. \ref{fig:network_architecture}. Table \ref{table:hyper-param-sr-cvae} shows the numbers of channels and strides in all convolution layers. The kernel size of all convolutions is 3. The slope of all leaky ReLU\cite{Maas+2013ProcICML} is $-0.01$. The upscale factor of all pixel shuffle operations\cite{Shi+2016CVPR} is 2. The resizing in the SR model is performed by bicubic interpolation. The weight parameters of all convolution layers are initialized randomly using uniform distributions.\cite{He+2015ICCV} For the Adam optimizer, the default parameters\cite{Kingma+2015ICLR} are used except for the learning rate of $1 \times 10^{-4}$. The mini-batch size is set to 128 for all experiments.

\begin{table}
    \caption{\label{table:hyper-param-sr-cvae} Hyperparameters for the convolution layers in the SR model and the CVAE. The layer names are defined in Fig. \ref{fig:network_architecture}.}
    \centering
    \begin{ruledtabular}
    \begin{tabular}{l r r r}
    Layer name & Input channels & Output channels & Stride \\
    \hline
    Conv2D-0 & 1 & 128 & 1\\
    Conv2D-1 & 128 & 128 & 1\\
    Conv2D-2 & 128 & 512 & 1\\
    Conv2D-3 & 128 & 256 & 1\\
    Conv2D-4 & 256 & 1024 & 1\\
    Conv2D-5 & 256 & 256 & 1\\
    Conv2D-6 & 256 & 1 & 1\\
    Conv2D-7 & 1 & 32 & 1\\
    Conv2D-8 & 256 & 1 & 1\\
    Conv2D-9 & 256 & 1 & 1\\
    Conv2D-10 & 1 & 8 & 1\\
    Conv2D-11 & 8 & 8 & 1\\
    Conv2D-12 & 8 & 1 & 1\\
    Conv2D-13 & 1 & 112 & 2\\
    Conv2D-14 & 112 & 112 & 1\\
    Conv2D-15 & 112 & 224 & 2\\
    Conv2D-16 & 224 & 224 & 1\\
    Conv2D-17 & 256 & 256 & 1\\
    \end{tabular}
    \end{ruledtabular}
\end{table}

\bibliography{reference}

\end{document}